# Gamma-ray Pulsar Revolution

Patrizia A. Caraveo

IASF-INAF

Via Bassini, 15

20133 Milano Italy

**Abstract**

Isolated Neutron Stars (INSs) were the first sources identified in the field of high-energy gamma-ray astronomy. At first, in the '70s, there were only two identified sources, the Crab and Vela pulsars. However, although few in number, these objects were crucial in establishing the very concept of a gamma-ray source. Moreover, they opened up significant discovery space both in the theoretical and phenomenological fronts. The need to explain the copious gamma-ray emission of these pulsars led to breakthrough developments in understanding the structure and physics of neutron star magnetospheres. In parallel, the 20-year-long chase to understand the nature of Geminga unveiled the existence of a radio-quiet, gamma-ray-emitting, INS, adding a new dimension to the INS family.

Today we are living through an extraordinary time of discovery. The current generation of gamma-ray detectors has vastly increased the population of known of gamma-ray-emitting neutron stars. The 100 mark was crossed in 2011 and we are now approaching 150. The gamma-ray-emitting neutron star population exhibits roughly equal numbers of radio-loud and radio-quiet young INSs, plus an astonishing, and unexpected, group of isolated and binary millisecond pulsars (MSPs). The number of MSPs is growing so rapidly that they are on their way to becoming the most numerous members of the family of gamma-ray-emitting Neutron Stars (NSs) .

Even as these findings have set the stage for a revolution in our understanding of gamma-ray-emitting neutron stars, long term monitoring of the gamma-ray sky has revealed evidence of flux variability in the Crab Nebula as well as in the pulsed emission from PSR J2021+4026, challenging a four-decade-old, constant-emission paradigm. Now we know that both pulsars and their nebulae can, indeed, display variable emission.

**Table of contents**

# 1-Introduction

Owing to their rapidly-rotating, hugely-intense magnetic fields, Neutron Stars are very efficient natural particle accelerators. Moreover, the accelerated particles (mostly electrons and positrons) are delivered into a highly-magnetized surrounding, ideal to make them radiate high-energy gamma rays that bear the timing signature of their parent INS. In a field such as high-energy gamma-ray astronomy, hampered by poor angular resolution compounded by relatively low number of detected photons, the presence of an unambiguous timing signature has been crucial to allow the identification of gamma-ray sources with crude positions . The combination of the ideal physical conditions with the telltale time signature makes INSs the most prominent class of high-energy gamma-ray emitters in our Galaxy.

However, progress in this field has been hampered by all sorts of experimental difficulties, stemming from the paucity of the gamma-ray photons. Gamma-ray pulsars cannot be detected in real time, like radio ones. To see pulsations, i.e. to build a statistically significant light curve, photons collected over anywhere from weeks to years have to be properly phased, folding their arrival times according to precise timing parameters. Prior to doing so, however, photon arrival times must be converted to the solar system barycenter, an ancillary yet inescapable operation upon which rests the success of the folding technique. Since such a correction is very position sensitive, a precise source position is an essential information for performing a search for pulsations in gamma rays . Thus, methods had to be devised to properly correct and fold the gamma-ray photon arrival times, while testing the statistical significance of the results.

In the meantime, the performance of gamma-ray telescopes has improved generation after generation going from NASA's pioneering SAS-2 (Fichtel et al., 1975) and ESA's COS-B (Bignami et al., 1975), to NASA's EGRET (Kanbach et al., 1988), to the current generation, encompassing ASI's AGILE (Tavani et al., 2009) and NASA's Fermi (née GLAST, Atwood et al., 2009). Four decades of unrelenting efforts in hardware and software were needed to go from the first firm detection of a gamma-ray pulsar to a family portrait with more than 100 objects. The growth, stemming primarily from the dramatic acceleration of recent years, is impressive, as shown in Figure 1. In parallel, the overall quality of the data, namely the angular, spectral and time resolution achieved for each photon as well as the overall sensitivity of the instruments, also significantly improved. This can be also seen in Figure 1, which shows a compilation of 5 sets of Vela pulsar light curves measured over a span of four decades by the five missions mentioned above.

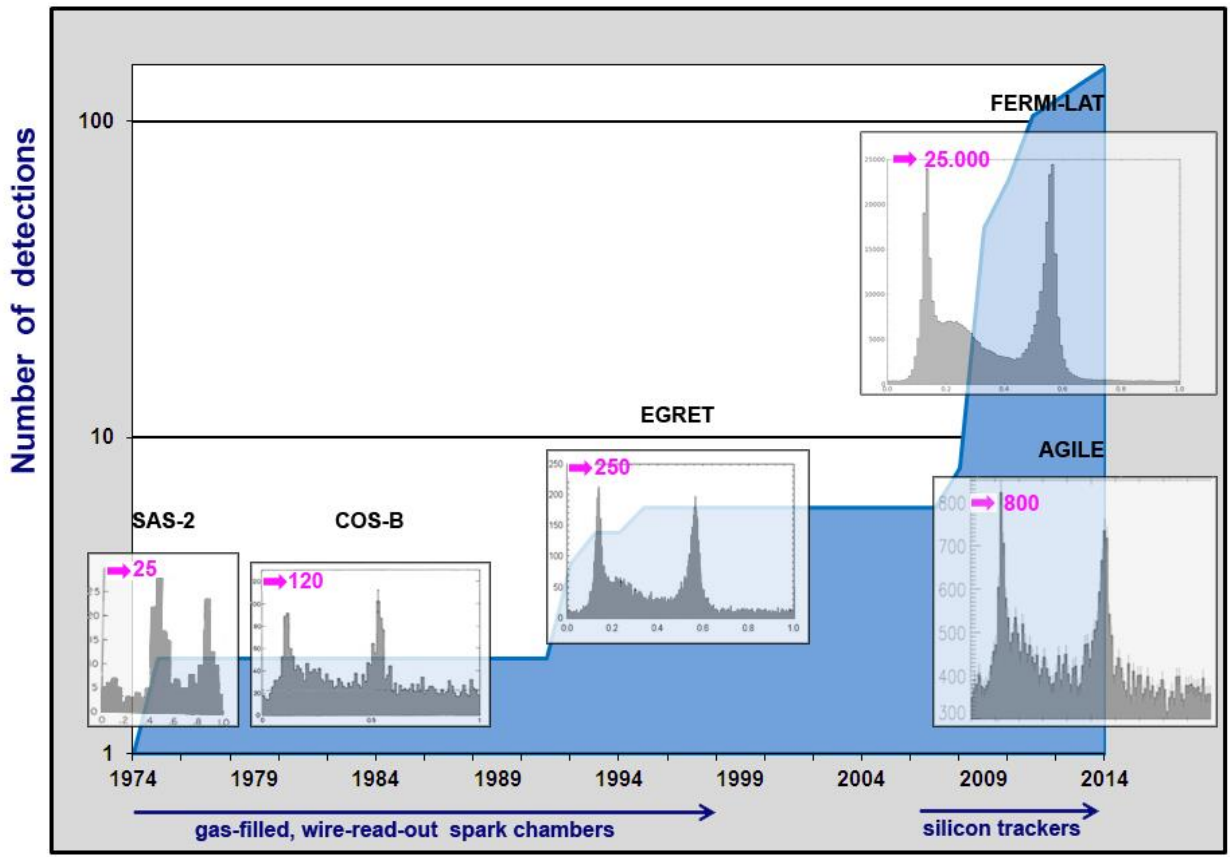

*Figure 1 Evidence of the beneficial effects of sensitivity increase. Over four decades of gamma-ray astronomy, going from gas-and-wire to silicon detectors, the total number of gamma-ray pulsars is seen to increase hundred folds (pale blue background area). The phase histograms (light curves) of gamma-ray photons collected from the Vela pulsar by SAS-2(>35 MeV), COS-B(>50 MeV), EGRET(>100 MeV), AGILE(>100 MeV) and Fermi LAT(>100 MeV) are also shown. The quality and detail level obtained (i.e. number of bins that can be afforded per graph and number of photons per bin) correlates directly with the increased photon statistics, i.e. mission sensitivity. The photon counting of the highest bin in the best light curve published by each mission is shown in magenta.*

## 2 The past : when sources were few and far between, and each one counted

**2.1 Pulsars as gamma-ray sources**

All the sky maps produced since the beginning of gamma-ray astronomy are dominated by three sources near the plane of the Galaxy. Folding the gamma photon arrival times using the radio ephemerides, SAS-2 identified first the Crab (Kniffen et al., 1974) and, later, the Vela pulsars (Thompson et al., 1975), i.e. the two brightest sources in the gamma-ray sky. The third source, shining next to the Crab in the Galactic anti-center, was named $\gamma$195+5 from its Galactic coordinates and could not be identified with any notable celestial object.

For the Crab, the SAS-2 detection was the confirmation of early, contradictory balloon claims (e.g. Vasseur et al., 1970, 1971), while for the older and less energetic Vela it was a genuine novelty. In the '70s pulsars were astronomical newcomers and the gamma-ray detections of Crab and Vela added an important new piece of information in the struggle to understand those extreme stars. The seminal work by Goldreich and Julian (1969) showed that a pulsar magnetospheres, far from being a vacuum, should be filled with plasma since the induced electric force wins over the gravitational pull on surface charges. Building on this result, Sturrock (1971) laid the foundations of pulsar electrodynamics and proposed the Polar Cap (PC) acceleration zone. Theoretical interpretations of the SAS-2 results elaborated on this idea, with contributions from Ruderman and Sutherland (1975) and Harding, Tademaru and Esposito (1978), further developed by Daugherty and Harding (1982). In PC models, particles, accelerated by rotation-induced electric fields above the polar cap, move along the dipole magnetic (B) field lines and produce curvature radiation. In this environment, photons above 1 GeV will be absorbed by the B field and produce $e^+ e^-$ pairs, which will radiate synchrotron photons and produce a second generation of pairs. Such a cascade will continue until the synchrotron photons fail to meet the energetic requirements to pair produce and can escape to contribute to the high-energy pulsar emission. The remaining pairs may supply particles to a coherent process that is responsible for the radio emission. This process takes place at low altitudes (< 1 $R_{NS}$, neutron star radius) above the stellar magnetic poles, where the beam of radio emission is also originating.

The premature termination of the SAS-2 mission interrupted its stream of discoveries, which was then taken up by COS-B, launched by the European Space Agency in 1975.

Similar to SAS-2 in dimensions (and, thus, in sensitivity), COS-B lasted much longer and had ample time to confirm and significantly improve the SAS-2 findings, while also discovering a population of unidentified Galactic gamma-ray sources (Swanenburg et al., 81).

A detailed study of the Crab and Vela behavior made it clear that the measured gamma-ray emission accounted for at least $10^{-3}$ of their rotation energy loss $\dot{E}$ ($\dot{E}_{rot} = -4\pi^2 I \dot{P} P^{-3}$ where P and $\dot{P}$ are the pulsar period and period derivative while I is the moment of inertia assumed to be $10^{45}$ g/cm$^2$). With a gamma-ray yield clearly dominating the pulsar emission, the gamma channel stands out as the most energetically demanding (Buccheri et al., 1978; Kanbach et al., 1980). To meet such requirements, Arons (1983) developed the Slot Gap (SG) model where pair creation takes place along favorably curved B-field lines above the polar caps and far from the Neutron Star (NS) surface. This model works well for short period pulsars, such as Crab and Vela. Although their double-peaked gamma-ray light curves appear similar, it was then immediately clear that their ratios between the pulsar rotational energy loss $\dot{E}$ and its gamma-ray luminosity $L_\gamma = 4\pi d^2 F_\gamma f_\Omega$ were different (where d is the pulsar distance, $F_\gamma$ the measured flux, and $f_\Omega$ is the beaming factor, which depends on pulsar geometry and is assumed to be 1 steradian, i.e. $f_\Omega = 1/4\pi$). The older and less energetic Vela was more efficient than the younger and more energetic Crab. The spectral shapes of the sources were different: while the Crab could be fitted with a single power law ($\frac{dN}{dE} = K \left(\frac{E}{E_0}\right)^{-2.1+/-0.3}$), for Vela a flattening at low energies combined with a steepening at high energies made a single power law fit much more challenging, if not impossible (Bennet et al., 1977). Moreover, their multi-wavelength behavior appeared vastly different. While the Crab exhibits similar light curves at all wavelengths, Vela's light curves are radically different at different wavelengths (e.g. Bignami and Hermsen, 1983). Such macroscopic effects, which could not be accounted for by the PC model, prompted Cheng, Ho and Ruderman (1986) to propose an outer magnetosphere model, where particles are accelerated within the vacuum outer gap extending from the null surface (i.e. where Ω·B = 0, which spatially separates the opposing charges) to the light cylinder (a virtual cylindrical surface or radius $R_{LC}$ where co-rotation stops, since $\Omega\, R_{LC} = c$), and produces gamma rays far from the neutron star surface, mainly by curvature and synchrotron radiation. The Outer Gap model (OG) applies only to gamma-ray emission and disentangles the gamma-emitting region from the radio one. The PC and OG models produce gamma rays in totally different regions of the pulsar magnetosphere, relatively near the surface for the polar cap scenario and far away for the outer gap, which approaches the light cylinder. The two different locations imply different ambient B fields, thus different emission-absorption combinations, as well as different geometry and thus different beaming. Moreover, "geographically" different emitting regions could come into view during the pulsar rotation, thus contributing at different pulsar phases to yield spectral shapes varying as a function of the pulsar rotation phase. The composite geometry interpretation was supported by the varying spectral shapes found in different phase intervals for the Vela pulsar by Grenier et al. (1988) in their attempt to perform phase-resolved spectroscopy.

**2.2 More gamma-ray sources: Geminga and the rest of the crowd….**

In addition to Crab and Vela, COS-B detected two dozen gamma-ray sources (Swanenburg et al., 1981) for which there were no obvious identifications. Due to COS-B's uneven coverage of the sky, such sources were mainly located near the Galactic plane, with the notable exception of an excess positionally coincident with 3C273, the first extragalactic gamma-ray source (Bignami et al., 1981a). Particular attention was devoted to γ195+5, readily dubbed Geminga (a name inspired by the gamma-ray source position within the GEMINi constellation, but also a pun in Milanese dialect meaning *is not there* or *there is nothing*).

Both SAS-2 (Ogelman et al., 1976) and COS-B (Buccheri et al., 1978) searched for pulsed signals from the known radio pulsars and both claimed low significance detections. While such results could not be confirmed, the statistical evaluation of the search outcome led to the development of the $Z_n^2$ method of gauging the statistical significance of a light curve (Buccheri et al.,1983) . De Jager et al. (1989), further elaborated in de Jager and Büsching (2010), proposed a different statistical test known as the H test. Both $Z_n^2$ and the H-test are now widely used.

However, no statistical test can overcome the irregularities usually present in pulsar timing behavior that hamper, and sometimes prevent, the use of archival, years-old, timing information to phase-fold the meager gamma-ray photon harvest. It was quickly realized that, if the radio timing parameters are not contemporary to the gamma-ray observations, the folding must be performed over a range of P and $\dot{P}$ values sampling the "extrapolated" parameter space. By multiplying the trials performed, this procedure weakens the significance of any tentative detection. Such a drawback was an important lesson learned for future instruments, and pointed to the need for contemporaneous radio monitoring of promising pulsars. In the meantime, the SAS-2 and COS-B claims, though unconfirmed, spurred the discussion on the evolution of gamma-ray emission efficiency as a function of pulsar age. If true, they would have implied a steady growth of the gamma-ray yield as pulsars age. While all those tentative detections have long been forgotten, the evolution of the gamma-ray emissivity as a function of pulsar age is still debated.

Meanwhile, by exploiting the latitude distribution of the unidentified sources (Swanenburg et al., 1981), as well as the shape of their LogN-LogS distribution (Bignami and Caraveo, 1980), it was possible to compute the average source luminosity and distance, showing that young, energetic pulsars at few kpc distance and with efficiencies between those of Crab and Vela could account for at least a fraction of the newly-found sources. This finding spurred the search for radio pulsars within the COS-B error boxes. However, the effort took some time and its results came too late to be useful to identify COS-B sources but paved the way for successful identifications with the following gamma-ray mission.

In parallel, a program to cover several COS-B unidentified sources error boxes by exploiting the imaging capability of the newly launched Einstein Observatory was successfully carried out as an alternative way to search for gamma-ray source counterparts (Caraveo, 1982). Much of the interest was focused on Geminga (Bignami et al. 1983), by far the brightest among the unidentified sources which had already defeated radio searches, but seminal results were obtained for 2CG135+01 (Bignami et al, 1981), for which the peculiar binary system LSI 61° 303 was proposed as a counterpart .

The chase for Geminga went on after the demise of COS-B, exploiting all the space and ground instruments available at all possible wavelengths, and bridging the hiatus between COS-B and the launch of EGRET. Bignami and Caraveo (1996) have summarized the long and checkered story which lead to the discovery of the first bona fide INS pulsating in X and gamma rays but not at radio wavelengths. Indeed, the gamma-ray pulsation was found only when ROSAT secured the X-ray periodicity detection (Holt and Halpern,1992) making it possible to fold the gamma-ray data collected almost simultaneously by EGRET (Bertsch et al., 1992) on board CGRO. Geminga had been pulsating all the time, of course, but the paucity of photons detected over COS-B's seven-year lifetime, together with the poor source localization, hampered the search for pulsations which could be found only a posteriori (Bignami and Caraveo, 1992). In retrospect, the discovery of a radio-quiet INS should not have been totally unexpected. In view of the geometry-driven emission mechanisms believed to be at work in radio pulsars, with radio-emitting regions probably detached from the gamma-ray regions, radio-quiet objects could have been expected. However, finding the first radio-quiet INS made it clear that gamma-ray astronomy had significant discovery space of its own.

**2.3 EGRET: more pulsars, at last**

Apart from studying the Crab (Nolan et al., 1993), Vela (Kanbach et al., 1994) and Geminga (Mattox et al., 1992, Mayer-Hasselwander et al., 1994) , EGRET, which could count on a significant contemporary pulsar radio monitoring campaign, detected three more pulsars, namely PSRB1706-44 (Thompson et al., 1992; 1996), PSRB1055-52 (Fierro et al., 1993) and PSRB1951+32 (Ramanamurthy et al., 1995), while PSRB1509-58 was only detected at low energy by the Comptel instrument (Kuiper et al., 1999), also on board Compton Gamma-Ray Observatory (Gehrels et al., 1994). We note that PSR B1706-44 coincides with the COS-B source 2CG342-02, but the radio pulsar was discovered after the end of the COS-B mission.

While Nel et al. (1996), using 3.5 yrs of EGRET observations, computed upper limits for 350 pulsars for which radio monitoring was available, it is worth mentioning that ad hoc searches did show that three

more pulsars, PSR B0656+14 (Ramanamurthy et al., 1996), PSR B1046-58 (Kaspi et al., 2000) and PSR J0218+4232 (Kuiper et al., 2000), were worthy of further investigation, although their detections were not yet compelling. While PSR B0656+14 and PSR B1046-58 were normal radio pulsars, similar to those already detected, PSRJ0218+4232 was an old recycled millisecond pulsar (MSP) characterized by extremely fast rotation coupled with a magnetic field significantly lower than that of normal pulsars. Harding et al. (2002, 2005) proposed a model for acceleration in the open field line region above the PC that seemed particularly well suited for MSPs.

Thompson (2004) provides a comprehensive review of the EGRET pulsar results. Ordering the known pulsars on the basis of their overall energy output $\dot{E}$ divided by their distance factor ($4\pi$ d$^2$), it became immediately clear that the pulsars so far detected in gamma rays were those ranking at the top of the list, i.e. those with the most favorable combination of energetics ($\dot{E} > 10^{34}$ erg/sec) and distance (Thompson et al., 1999). Although it does not account for important variables, such as different inclinations and viewing angles as well as different efficiencies, and its value is only as good as the (usually uncertain) distance estimate, the $\frac{\dot{E}}{4\pi d^2}$ ranking (and its variant, $\frac{\sqrt{\dot{E}}}{4\pi d^2}$) proved to be an extremely useful tool.

Plotting the pulsars' gamma-ray luminosities $L_\gamma$ as a function of the open field line voltage à la Goldreich and Julian (1969) a trend is seen pointing to a proportionality between $L_\gamma$ and the open field line voltage, itself proportional to $\sqrt{\dot{E}}$ (Thompson, 2004).

All pulsars exhibit a power law spectral shape with a high-energy cut off. Light curves are usually double peaked, and the peak ratio varies with energy, with the second peak usually harder than the first (by definition, the "first" gamma-ray peak is the one that comes in phase immediately after the main radio peak). Reproducing such a double-peaked structure turns out to be an important test for pulsar models. Chiang and Romani (1992,1994) argued that the EGRET pulsars' light curves, as well as their spectra, arise naturally from a modified version of the OG model. Working in 3D and accounting for both the angle $\zeta$ between the observer's line of sight and the NS rotation axis, and the dipole field inclination angle $\alpha$ (see the scheme of Figure 2), Romani and Yadigaroglu (1995) succeeded in reproducing the Vela light curve as measured at radio, optical, X-ray and gamma-ray wavelengths for a very inclined geometric combination of α=65° and ζ=90°. Indeed, to produce double peaked light curves, OG models do prefer highly inclined rotators (see also Romani, 1996), while PC models require nearly aligned geometry, allowing magnetic inclination angles comparable to the angular extent of the polar cap. To overcome the PC requirements on pulsar alignment, Daugherty and Harding (1996), elaborating on the seminal work by Arons (1983) on the Slot Gap idea, extrapolated the polar cap acceleration region to higher altitudes, leading to a full SG model by Muslinov and Harding (2003,2004). Here particle acceleration takes place in thin slot gaps along the last open field line connecting the neutron star surface to the light cylinder. By extending the acceleration, and thus the gamma-ray production region, the SG model can be adjusted to any magnetic inclination angle. Including special relativity effects, such as aberration and time of flight delay, in the SG framework, Dyks and Rudak (2003) developed a Two-Pole Caustic model (TPC) that, in their opinion, can overcome some of the shortcomings of both PC and OG. To add freedom to the OG model, Hirotani et al. (2003) proposed an outer gap that extends beyond the null surface. Production inside the light cylinder, however popular, is not the only option for pulsar modeling: Coroniti (1990) proposed a totally different approach with the striped wind model where gamma rays are produced outside the light cylinder.

Apart from fitting light curves and spectra, models must account for the measured efficiency in converting rotational energy into gamma rays since a pulsar's gamma-ray yield is by far the dominant component of its multi-wavelength emission. Romani (1996) as well as Arons (1996) discuss how to get high yields from OG and SG models. However, no model could account for a gamma-ray luminosity exceeding $\dot{E}$, as seemed to be the case for PSR B1055-52 (Thompson et al., 1999). That pulsar's gamma-ray luminosity, computed assuming a beaming factor of 1 steradian, and using the distance derived from its radio dispersion measure (DM), required an efficiency of more than 100%, pointing to a wrong distance estimate, and possibly, too large a beaming factor. Clearly, coupling a few % "true" pulsar efficiency with an overestimated distance and an uncertain beaming could produce an unreasonably high gamma-ray yield.

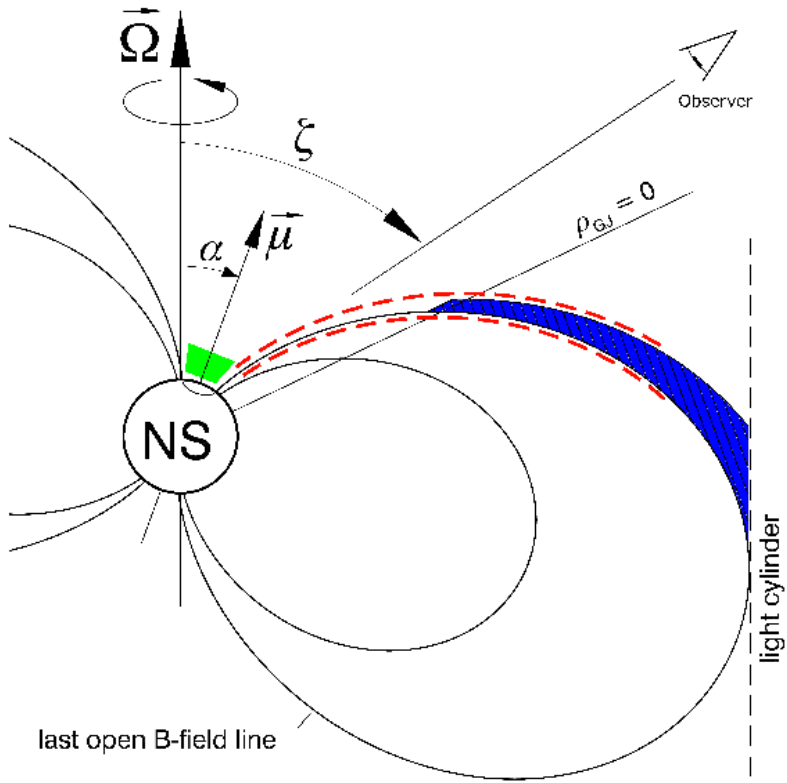

*FIGURE 2: Scheme of a NS magnetosphere with the internal emission regions highlighted: PC is in green, OG in dark blue and SG in red (from Dyks and Rudak, 2003). The striped wind scenario (outside the light cylinder) is not shown.*

Since the light curves of all the newly discovered EGRET pulsars appeared to be different at different wavelengths and generally not aligned in phase (see e.g. Thompson, 2004), thus following the Vela paradigm as opposed to the Crab one, the need to disentangle the radio emission region (almost certainly related to the polar cap) from the gamma-ray region certainly supported OG or SG emission models rather than the classical PC model.

However, while "near" and "far" emission models could be adjusted to produce the double-peaked light curves seen by EGRET, they differed in the gamma-ray spectral shape, with the polar cap model unable to produce very high-energy photons. Because of the absorption in high-B fields, the PC model predicts sharp, super-exponential cutoffs in the observed spectra at energies of ∼ few GeV. However, high-altitude models, such as OG and SG, predict a simple exponential cutoff due to the radiation reaction limit of the accelerated particles. The EGRET sensitivity above a few GeV, unfortunately, did not allow discrimination between the two classes of models.

**2.4 EGRET : hundreds of gamma-ray sources and many interesting candidates**

Although EGRET had increased the number of gamma-ray emitting INSs, pulsars were no longer the dominant celestial population in the gamma-ray sky. Extragalactic sources, many of them strongly variable, were now counted by the dozens. They greatly outnumbered the INSs in the final EGRET catalogue, which listed 278 sources (Hartman et al., 1999), half of which remained unidentified. However, with 2/3 of the unidentified sources clustered around the Galactic plane, pulsars, both radio-loud and radio-quiet, continued to be very natural candidates to account for non-variable sources with no obvious counterpart. Repeating the geometric exercise with the EGRET low-latitude sources, Mukherjee et al. (1995), confirmed the findings of Swanenburg et al. (1981). The unidentified low latitude EGRET sources lay at distances between 1.2 and 6 kpc and their luminosities range from $0.7 \times 10^{35}$ to $16.7 \times 10^{35}$ erg s-1, values calling for rather young and energetic INSs. Gehrels et al. (2000) elaborated on the LogN-LogS distribution of the unidentified EGRET sources to claim the existence of a population of faint gamma-ray sources at mid-latitude .

While Geminga-like INSs were ideal, but elusive, potential counterparts (e.g. Yadigaroglu and Romani, 1995), deep radio searches started to detect promising young and energetic pulsars within EGRET error boxes. However, finding a radio pulsar in a gamma-ray error box does not guarantee the detection of pulsations in gamma rays . As mentioned above, if the gamma-ray observations are separated in time from the radio ones, the pulsar's period and period derivative must be extrapolated while also accounting for their uncertainties, so that the periodicity search in gamma rays must cover a vast parameter space, reducing the significance of any tentative detection. Moreover, young pulsars often exhibit timing noise and glitches which render the extrapolation much more uncertain. Thus, it should come as no surprise that the promising pulsars discovered at the end of the EGRET mission could not qualify for a solid detection of gamma-ray pulsations. As already mentioned, Kaspi et al. (2000) reported evidence of an association between the 20,000 yr-old PSR B1046-58 and 3EG J1048-5840, adding this pulsar to the interesting candidates list. D'Amico et al. (2001) using Parkes data, found two young, promising radio pulsars inside the error boxes of 3EG J1420-6038 and 3EG J1837-0606. More young and energetic pulsars were found while exploring EGRET error boxes: Roberts et al. (2002) uncovered PSRJ2021+3615 within the source 3EGJ2021+3716, in the rich and complex Cygnus region. Yet, the pulsar timing noise prevented a meaningful back-extrapolation searching for pulsations in the limited number of gamma-ray photons EGRET had collected. Similarly, PSR J2229+6114 is a young radio and X-ray pulsar whose energetics makes it a plausible counterpart for 3EG 2227+6122, within which it was found, though its dispersion measure pointed to a distance value in excess of 10 kpc. If true, such a distance would have made PSR J2229+6114 too faint to be the gamma-ray source counterpart (Halpern et al., 2001). Not surprisingly, a search for pulsations in the gamma-ray data was inconclusive.

In more general terms, by cross-correlating 1,300 known radio pulsars with the EGRET catalog, Kramer et al. (2003) estimated that 19±6 radio pulsar associations could have been genuine.

While the new young and energetic pulsars went into the wish list of high-energy astrophysicists, waiting for the next generation of gamma-ray instruments, these examples illustrate the complex interplay between gamma-ray and radio astronomy. Indeed, the discovery of radio pulsars within gamma-ray error boxes is a story that shall recur in our narrative, growing as time goes by.

Meanwhile, a multi-wavelength approach, exploiting the sequence of X-ray and optical observations that had been successfully applied to Geminga, was pursued for a number of bright, unidentified sources at medium to low Galactic latitudes. Notable examples are 3EG J1835+5918, dubbed "Next Geminga" owing to its similarities with the prototype radio-quiet INS. Featuring an X-ray counterpart detected by Chandra, RX J1836.2+5925, with both thermal and non-thermal emission, but without optical and radio detection, the steady EGRET source with a hard spectrum and high-energy cut off looked like a radio-quiet INS (Mirabal & Halpern, 2001; Reimer et al., 2001; Halpern et al., 2002).

Similarly, 3EG J2020+4017 (Brazier et al, 1996) and 3EG J0010+7309 (Brazier et al., 1998) appeared to be positionally associated with the supernova remnants gamma Cygni and CTA-1, respectively. The two gamma-ray sources are non-variable and have flat spectra similar to other gamma-ray pulsars. 3EG J0010+7309 has an X-ray counterpart that, again, looks like the young INS responsible for CTA-1, embedded in its plerion (Halpern et al., 2004).

Although now identified, Geminga continued to attract attention, becoming one of the most scrutinized INSs in the soft X-ray domain. Both ESA's XMM-Newton and NASA's Chandra X-ray observatories devoted significant observing time to this source. A long XMM-Newton observation unveiled a nebula trailing the neutron star as it moves in the interstellar medium (Caraveo et al., 2003). The same data set also allowed for phase-resolved spectroscopy to be performed, making it possible to disentangle the non-thermal, power-law component from the surface thermal emission, which was divided into hot and cool components that were seen to vary as a function of the pulsar rotational phase (Caraveo et al., 2004, de Luca et al., 2005). Using the precise knowledge of the distance to convert X-ray fluxes into luminosities, the emitting areas were computed, showing that the hot component (probably linked to the polar cap heating by return current) comes from a surface much smaller than that of any dipole-like polar cap. This pointed to a quasi-aligned rotator almost perpendicular to the line of sight. The nebula detected by XMM-Newton was confirmed by Chandra (Pavlov et al., 2010), which also resolved a comet-like structure trailing the pulsar (de Luca et al., 2006).

**2.5 The EGRET Legacy: Open Points for a new Millennium**

When CGRO was deorbited in June 2000, the EGRET mission legacy in pulsar astronomy amounted to as many as 10 INSs (seven firmly established detections and three probable ones). Although the objects had somewhat different phenomenology, they undoubtedly channeled the major share of their rotational energy loss in gamma rays. As shown by the energy-per-decade plot (Thompson, 2004) all the EGRET pulsar spectral energy distributions peak in the gamma-ray band . Their high efficiencies, coupled with extremely diverse multi-frequency behavior, clearly points to composite emission models where different emitting regions at different locations in the pulsar magnetosphere contribute all at once. However, while the various models could account for gamma-ray emission below 1 GeV, the Rosetta stone of pulsar modeling lay in the few to 10 GeV region where different models predicted very different spectral shapes. Since the Polar Cap model predicts a sharp turnover at few to several GeV (as a result of the attenuation of the gamma-ray flux in the magnetic field above the star surface), the lack of such a turnover would rule out the already troubled polar cap model in favor of the outer gap or slot gap, which produce gamma rays far from the pulsar's surface. In particular, the expected sensitivity of the Large Area Telescope (LAT), on board NASA's GLAST mission, to GeV photons gave the hope of solving this spectral conundrum, making it possible for scientists to pin down the best emission model (as shown by Razzano et al. (2009) using simulated LAT data).

EGRET had also left a number of promising unidentified sources for which a direct search for periodicity was beyond reach, but that could be investigated at other wavelengths following the lessons learned in the chase for Geminga. Indeed, radio and X-ray searches were being actively pursued, yielding a growing number of potential gamma-ray source counterparts that were just waiting for the next gamma-ray observatories.

Expectations of pulsar detections for the coming AGILE and GLAST missions were rather uncertain, mainly due to considerable uncertainty of which pulsar model to use. Back extrapolating the LogN-LogS number-flux relation of the EGRET pulsars, Thompson (2004) predicted a grand total ranging from 30 to 100 GLAST detections, considering both radio-quiet and radio-loud INSs. On the basis of Monte Carlo simulations for polar cap emitting pulsars, Harding et al. (2002) predicted the gamma-ray detection of 90 radio-loud and 101 radio-quiet pulsars. However, less than 10% of the radio-quiet pulsars were expected to yield a pulsed signal with the techniques available at the time. Simulations including outer gap emission models (Harding et al., 2007) yielded a large number of radio-quiet gamma-ray pulsars coupled to a comparatively small number of radio-loud ones, making it clear that the ratio of radio-quiet to radio-loud pulsars detected by GLAST would provide a useful clue in discriminating between the two classes of models. Moreover, Harding et al. (2005), on the basis of their PC model, predicted that several MSPs could be detectable by AGILE and GLAST, both yet to be launched.

Building on the lessons learned during past missions, careful campaigns involving all the major radio observatories were orchestrated to maximize the chances of success for detection of radio pulsars. Smith et al. (2008) describe the observing campaign organized by the GLAST collaboration together with all the major radio observatories (now known as the Pulsar Timing Consortium) to monitor about 200 pulsars selected for their large spin down power ($\dot{E}$ >10$^{34}$ erg/sec) and ranked high according to their $\frac{\sqrt{\dot{E}}}{4\pi d^2}$. This effort also included careful calibration of the spacecraft clock (Smith and Thompson, 2009; Abdo et al 2009) to allow for precise phase alignment of the radio and gamma-ray light curves. This rendered possible phase-folding of millisecond pulsars, a class of neutron stars often in binary systems, certainly interesting, but not considered a prime target for the mission, in view of their rather low surface B field. For the radio-quiet Geminga, Jackson & Halpern (2005) undertook the task of maintaining phase-coherent timing parameters through biannual XMM-Newton measurements.

Preparation efforts were also carried out in the software domain. Extensive "data challenges", relying on massive Monte Carlo simulations (Baldini et al., 2006) of the gamma-ray sky, were conducted within the GLAST collaboration to test and debug the analysis software, with the goal of having a fully functional analysis pipeline ready at the time of launch.

Meanwhile, to meet the challenge of detecting pulsations using only the gamma-ray photons, Atwood et al. (2006) devised a new strategy to optimize the computing power needed to perform comprehensive blind searches covering a wide range of P and $\dot{P}$. Since very long exposure times, which are mandatory in gamma-ray astronomy, make complete Fourier analysis computationally prohibitive, the newly proposed method analyzes the differences of photon arrival times (up to a window of the order of weeks) rather than the time series itself and succeeds in maintaining good sensitivity while greatly reducing the effects of frequency derivatives and glitches. The time-difference method was successfully tried on EGRET data (Ziegler et al., 2008), and was ready to be tested on new gamma-ray data.

## 3.The Present: the Silicon Era in gamma-ray astronomy

The new era in gamma-ray astronomy is characterized by a significant hardware improvement due to the introduction of the silicon-strip detector as the core constituent of a "solid-state" spark chamber (as opposed to the gas-filled, wire-read-out spark-chambers on-board SAS-2, COS-B and EGRET) used to detect and track photons and to discriminate between them and the much more numerous charged particles. Silicon trackers are self-triggering devices that reduce the dead time to almost zero, enhancing the instrument's timing and count-rate capabilities while also improving its spectral and spatial resolution. Moreover, their very compact structure allows for the design of gamma-ray detectors with steradian-size Fields of View (FoV), thus increasing the effective observing time for any given direction in the sky.

Currently (October 2013), we live in an ideal time for high-energy astrophysics, with two such instruments operating in orbit allowing, for the first time since the beginning of gamma-ray astronomy, almost continuous coverage of the sky, plus the possibility to independently prove (or disprove) each other's claims.

The two satellites are AGILE (Tavani et al., 2009), a small Italian mission of the Italian Space Agency (ASI) in close collaboration with the Istituto Nazionale di Astrofisica (INAF) and the Istituto Nazionale di Fisica Nucleare (INFN), and *Fermi* (the name given to GLAST in orbit, Atwood et al, 2009), a much bigger NASA mission with important international participations from Japan, France, Italy (again with ASI, INAF and INFN) and Sweden.

**3.1 April 2007: enters AGILE**

AGILE features a silicon tracker made of 12 40x40 cm trays and a thin calorimeter. The mass of the calorimeter is a limitation for the detection of photons with energies > tens of GeV, but AGILE excels at low energies (E<100 MeV). Above the tracker, another layer of silicon acts as a hard X-ray detector and is known as SuperAGILE. Although the dimensions of AGILE are about a quarter of those of EGRET, their effective areas are comparable thanks to the superior performance of the silicon tracker and analog read-out, with AGILE enjoying significantly better angular and time resolution, as well as a much smaller dead time.

Launched in April 2007 from the Indian base of Sriharikota, AGILE was put into a nearly equatorial orbit and started its observing program as a pointing mission characterized by a very large FoV, covering one sixth of the sky. As usual, AGILE carried out its "on-orbit" calibration using the Vela and Crab pulsars as targets, and also testing its capabilities with Geminga and PSR B1706-44 (Pellizzoni et al., 2009a). After subtracting the pulsed photons from the Vela source, a faint diffuse emission appeared, yielding the first clear detection of a resolved Pulsar Wind Nebula (PWN) in high-energy gamma rays (Pellizzoni et al., 2010).

The first new gamma-ray pulsar was discovered as part of the AGILE Guest Observer Program by Halpern et al. (2008), who detected gamma-ray emission from PSR J2021+3651, showing that its radio distance was indeed overestimated. Thus, PSR J2021+3651 entered the stage as one of the contributors of the notoriously complex Cygnus region. "*Next Geminga*" was also closely scrutinized in order to ascertain its spectral shape as well as search for possible variability (Bulgarelli et al., 2008)

Folding the gamma-ray photon arrival times using contemporary ephemerides for the 35 top-ranking radio pulsars (for which radio monitoring had been organized), AGILE detected three new pulsars (Pellizzoni et al., 2009b), namely PSR J2229 + 6114, discovered by Halpern et al. (2001) within the 3EG 2227+6122 error box, PSR B1509-58, the very energetic pulsar that had been seen by COMPTEL but not by EGRET owing to its very soft spectrum, and PSR J1824-2452, a MSP detected clearly for only a fraction of the observing time. PSR B1509-58 was further investigated by Pilia et al.(2010). Moreover, Pellizzoni et al. (2009b) found tantalizing signals for four more objects, namely PSRs J1016–5857, J1357–6429, J2043+2740, and J1524–5625. Of these, only the latter, which was also the least significant, was not independently detected by *Fermi*.

By doubling the EGRET pulsar harvest, adding the detection of the second youngest as well as the million-year old PSR J2043+2740, AGILE was showing that gamma-ray emission is a common feature of high ranking radio-loud pulsars, be they young or old.

**3.2 June 2008: enters Fermi and finds a new Geminga**

In June 2008 GLAST was launched and, thus, the *Fermi* observatory was born featuring two major instruments: the Large Area Telescope (LAT, Atwood et al, 2009) and the Gamma Burst Monitor (GBM, Meegan et al, 2009). The LAT modular structure, containing 16 towers (each with dimensions comparable to AGILE), featuring an 18-tray tracker sitting on a heavy calorimeter , makes it the most powerful gamma-ray telescope ever. Very well suited to detecting photons in the GeV range, where angular resolution is at its best, Fermi aimed at obtaining the sharpest (and deepest) vision of the gamma-ray sky. With a FoV of 2.4 steradian, Fermi was designed to primarily operate in scanning mode, covering the whole sky every three hours, i.e. every two 90 min. orbits. Such an operating mode provides a reasonably homogeneous sky coverage and guarantees that any given point in the sky is within the instrument FoV for ~one sixth of the time. Once in orbit, Fermi executed a planned calibration sequence of pointed observations during which the satellite pointed first at the Vela pulsar and, when the target was occulted by the Earth, at the EGRET unidentified source 3EG J0010+7309, coincident with the CTA-1 supernova remnant.

Thanks to the instrument performance, the software readiness, the differencing technique and the availability of a precise X-ray position, *Fermi's* On-Orbit verification phase yielded the long-sought detection of periodicity from 3EG J0010+7309, making it the first high visibility result (Abdo et al., 2008) of the new *Fermi* observatory. While 3EG J0010+7309 was quite a robust INS candidate, with a faint X-ray source surrounded by diffuse emission pointing to a neutron star embedded in a PWN (Halpern et al., 2004), the speed of the discovery was amazing, boding well for the future capability of *Fermi* as a hunter of radio-quiet pulsars (Bignami, 2008).

The pulsar timing parameters yielded a characteristic age ($\tau_c = \frac{P}{2\dot{P}}$) of 14,000 years (comparable to the CTA-1 SNR age) and a rotational energy loss of $4.5 \times 10^{35}$ erg s$^{-1}$, i.e. a radio-quiet INS 50 times younger and 10 times more energetic than Geminga. Its dipole B field is rather high $1.1 \times 10^{13}$ G, which makes PRS J0007+7303 the second-highest magnetic field pulsar after PSR B1509-58. However, unlike PSR B1509-58, the newly discovered INS does emit high-energy photons. The pulsar light curve displays a double-peaked structure similar to that of PSR B1706-44, one of the historical EGRET pulsars that has a similar age. For radio-quiet pulsars, distance estimates rest on indirect methods such as X-ray line-of-sight absorption or their association with a SNR. Combining the radio and X-ray pieces of information, the best distance to PSR J0007+7303 is $1.4 \pm 0.3$ kpc (Pineault et al, 1993). This makes it possible to compute the pulsar luminosity - and thus its efficiency - that turns out to be about 1%. Interestingly, the efficiency of PSR J0007+7303 in converting its rotational energy loss into gamma rays is similar to that of PSR B1706-44, but smaller than that of the much older Geminga. Although PSR J0007+7303 is a Geminga-like gamma-ray pulsar with an X-ray counterpart, we note that, unlike Geminga, here, the gamma rays were the driver behind the periodicity discovery. The source faintness in X-rays prevented the detection of pulsations in that band. Only a long XMM-Newton observation, coupled with the precise knowledge of the gamma-ray timing parameters, yielded the X-ray pulsation detection (Caraveo et al., 2010). The X-ray counterpart of PSR J0007+7303 is a faint source, excellent for getting an accurate position, but of hardly any use for unveiling a hidden periodicity.

Pulsations from the radio pulsar PSR J1028-5819, freshly discovered by Keith et al. (2008) within 3EGJ1027-5817, followed quickly (Abdo et al 2009a ), identifying the pulsar as being responsible for at least part of the flux of 3EG J1027-5817. Next came the detection of gamma-ray pulsations from PSR J0205+6449 in 3C58, a Crab-like 800-year old SNR (Abdo et al., 2009b). The pulsar light curve is also Crab-like and its two peaks are aligned with the X-ray ones (but not with the single radio pulse).

An impressive light curve of the Vela pulsar, including 32,400 pulsed photons collected during the verification phase, was quickly published (Abdo et al 2009c) with a revealing spectral study of the phase-averaged gamma-ray emission. The shape of the light curve as a function of energy confirms the dramatic energy evolution hinted at by previous experiments with the appearance of a third peak in the bridge region above 1 GeV and the disappearance of the first peak at energies above 10 GeV. The *Fermi* spectrum can be described as a power law with exponential cutoff $\frac{dN}{dE} = K \left(\frac{E}{E_0}\right)^{-\Gamma} exp\left(-\frac{E}{E_{cut}}\right)^b$ with spectral index $\Gamma$ = 1.5, an energy cutoff, $E_{cut}$=2.9 GeV and b of 0.88 ± 0.4. This rules out any super-exponential absorption that would have been the signature of attenuation due to interaction of high-energy gamma-ray photons with the strong magnetic field at low-altitude in the pulsar magnetosphere. Thus, the lack of hyper exponential absorption points to high-latitude emission. This finding is strengthened by the detection of pulsed photons up to 17 GeV, an emission that must arise at R > 3.8 $R_{NS}$ (from Baring, 2004).

The first high-quality *Fermi* pulsar spectrum set the stage for gamma-ray emission far from the NS surface in the outer magnetosphere, near the light cylinder, ruling out the PC model in favor of the outer gap or slot gap models.

**3.3 Surprises: expected and unexpected**

Whereas gamma-ray emission from young, energetic pulsars (both radio-loud and radio-quiet) was widely expected, the detection of the millisecond pulsar PSR J0030+0451 (Abdo et al., 2009d) came as a real surprise, mainly because such old, recycled pulsars were not supposed to be ideal gamma-ray emitters as their surface B fields are $10^4$ times weaker than those of young NSs.

Conversely, the pulsar gamma-ray phenomenology was pretty normal looking: two narrow peaks, separated by 0.44 in phase, a spectrum well-described by a power law with an exponential cut off. However, its rather low rotational energy loss of 3.5x$10^{33}$ erg s$^{-1}$, coupled with the parallactic distance of 300 pc and a the measured flux (E>100 MeV) of approximately 7 $10^{-8}$ cm$^2$ sec$^{-1}$, implied a huge 15% gamma-ray production efficiency.

The detection of PSR J0030+0451 opened the way for a systematic search for gamma-ray emission from MSPs. Owing to the radio monitoring campaign (Smith et al., 2008), MSPs with $\dot{E}$ above $10^{34}$ erg/sec had contemporary ephemerides to be used for phase folding. The search immediately yielded interesting results with 7 more MSPs found in a matter of months (Abdo et al., 2009e). Although MSPs are often in binary systems, we will continue to refer to gamma-ray pulsars as Isolated Neutron Stars, since their emission is powered only by their rotational energy loss. With gamma-ray light curves and spectra similar to those of young pulsars, the emission regions for MSPs should also have been far from the neutron star surface. Although young pulsars and millisecond ones have vastly different B-fields at the stellar surface, the value of the B field at the light cylinder is similar , indicating a region where similar conditions naturally occur.

All the eight gamma ray emitting MSPs were field objects, both isolated and in binary systems. The detection of the globular cluster 47 Tuc (Abdo et al., 2009f ) was therefore attributed to the integrated emission of multiple MSPs within it. The case of 47 Tuc is not unique; *Fermi* has detected a dozen more globular clusters as point sources (Abdo et al., 2010a, Kong et al. 2010). In parallel, blind searches over a five-month observing interval yielded 15 more radio-quiet INSs, making it clear that radio-loud and radio-quiet were evenly contributing to the *Fermi* pulsar harvest. Abdo et al. (2009g) discuss the general properties of the population of 16 radio-quiet INSs discovered by *Fermi*. Since 13 of them were found within formerly unidentified EGRET sources that were among the set of suspected pulsars, deep X-ray investigations were already available. Indeed, six of the sixteen pulsars were discovered by assuming a counterpart position derived from previous X-ray coverage or from newly-obtained observations, mainly using the Swift X-ray Telescope. Apart from CTA-1, the newly –discovered gamma-ray pulsars featured: "Next Geminga" (Halpern et al., 2004), "Gamma Cyg" (Brazier et al., 1996), the "Rabbit" (Ng et al., 2005), "Taz" (Roberts et al., 2008), and the "Eel" (Roberts et al., 2001). However, the faintness of the X-ray counterparts made it clear that, for gamma-ray pulsars, the LAT's potential as a pulsation discoverer was vastly superior to that of X-ray telescopes.

The timing parameters of the 16 gamma-ray selected pulsars point to rather young INSs, with energetics similar to that of radio-loud ones. Also, their averaged fluxes, light-curves and spectra were no different than that of radio-loud pulsars. Subsequent deep radio observations yielded detections for three of the new pulsars; two rather normal looking and one exceedingly faint (Camilo et al., 2009, Abdo et al., 2010h).

A closer look at bright sources such as Vela (Abdo et al., 2010e), the Crab (Abdo et al., 2010c), Geminga (Abdo et al., 2010f) as well as "Next Geminga" (Abdo et al., 2010d), for which phase-resolved analysis was possible, did clearly showed that both spectral index and cutoff energy were varying as a function of the pulsar phase, confirming early findings and adding a wealth of fine structure details. The detection of the Crab pulsar at E > 100 GeV by Magic and Veritas (Aliu et al., 2008, 2011; Aleksic et al., 2012) is well above any reasonable extrapolation of the *Fermi*-LAT best spectral fit, pointing to a different emission mechanism for that high-energy component.

The first *Fermi* catalog of gamma-ray pulsars (1PC) lists 46 high-confidence pulsars detected within the first 6 months of the mission (Abdo et al., 2010b). Out of 46 NSs, 29 were detected in radio (further divided between 8 MSPs and 21 young pulsars) and 17 seen only in gamma rays (i.e. 16 discovered by LAT + Geminga). 15 of the newly discovered LAT pulsars had been found within the error boxes of high-interest EGRET unidentified sources, confirming the pre-launch expectations. 13 of these were radio-quiet and 2 were radio-loud, namely PSR J1028-5918 (Abdo et al., 2009a) and PSR J2021+3651 (Abdo et al., 2009h).

A power law with and exponential cutoff can fit the spectra of all 46 gamma-ray pulsars. Their light curves are usually double peaked (with peak separation of 0.4-0.6), but a non-negligible minority of single-peaked pulsars is also present. With very few exceptions, the gamma-ray peaks are not aligned with the radio ones, confirming early EGRET findings and pointing to an emission region far from the pulsar's surface.

To assess the luminosity of gamma-ray pulsars ($L_\gamma = 4\pi d^2 F_\gamma f_\Omega$) a $f_\Omega = 1$ was used for all pulsars. It is a big change compared to the past, when this parameter was assumed to be $1/4\pi$, and is a direct consequence of the new preference for the outer gap model (see, e.g. Watters et al., 2009). The gamma-ray light curves and spectral shapes point to high-altitude emission regions producing fan beams that cover a large fraction of the celestial sphere. However, when computing the luminosity of LAT pulsars, the major source of uncertainty remains the assumed distance , since few INSs have a measured parallax. While the majority of radio pulsars can rely on dispersion measure, distances for radio-quiet pulsars rest only on X-ray absorption, when it is available. This limits the number of radio-quiet pulsars in the *Fermi* luminosity plot.

Abdo et al. (2010b) show that the evolution of the gamma-ray luminosity as a function of the pulsar rotational energy loss cannot be fitted by a single function. Even considering the distance uncertainties, a substantial scatter is present, possibly arguing against the assumption of a common beaming factor for all objects. MSPs seem to climb more steeply in luminosity than the young pulsars, which evolve more gently.

**3.4 *Fermi*'s Treasure Hunt**

Numerous multi-wavelength studies were triggered by the first wave of *Fermi* results, both directly linked to pulsars and, more generally, to newly detected sources with no identification, first in the Fermi Bright Source Catalog (often referred to as 0FGL Abdo et al., 2009i), and later in the First List of *Fermi* Sources, known as 1FGL (Abdo et al, 2010g). On the high-energy side, both exploratory and in-depth X-ray observations were carried out targeting the newly-discovered radio-quiet INSs as well as the promising candidates that were emerging from the *Fermi* data, but that needed better position information to secure a statistically significant pulsation detection. On the low-energy side, the time-honored exercise to search for radio pulsars within unidentified source error boxes was started anew. By exploiting the smaller *Fermi* error boxes, as well as more powerful analysis systems, it was possible to significantly reduce the time needed to cover each source. This allowed for multiple visits, a strategy that proved crucial to detecting radio emission from MSPs in binary systems. To maximize the chances of success, sources to be studied were selected on the basis of their "*pulsarness*", a parameter quantifying the lack of variability coupled to a suitably curved spectral shape (Ackermann et al., 2012).

Having searched 25 such pulsar-like unidentified sources, Ransom et al. (2011) reported the detection of 3 new MSPs. Such a high success rate triggered the chase for field MSPs in unidentified pulsar-like *Fermi* sources, preferably at medium to high Galactic latitude. Keith et al. (2011) found 2 MSPs and a young pulsar in their coverage of 11 *Fermi* sources. Cognard et al. (2011) unveiled two MSPs, while Kerr et al. (2012) found 5 MSPs millisecond pulsars after having searched 14 unidentified *Fermi* sources. Once the timing parameters of the often-binary new MSPs were accurate enough, they were also detected in gamma rays, increasing the share of this class of rather faint gamma-ray emitters (both isolated and in binary systems).

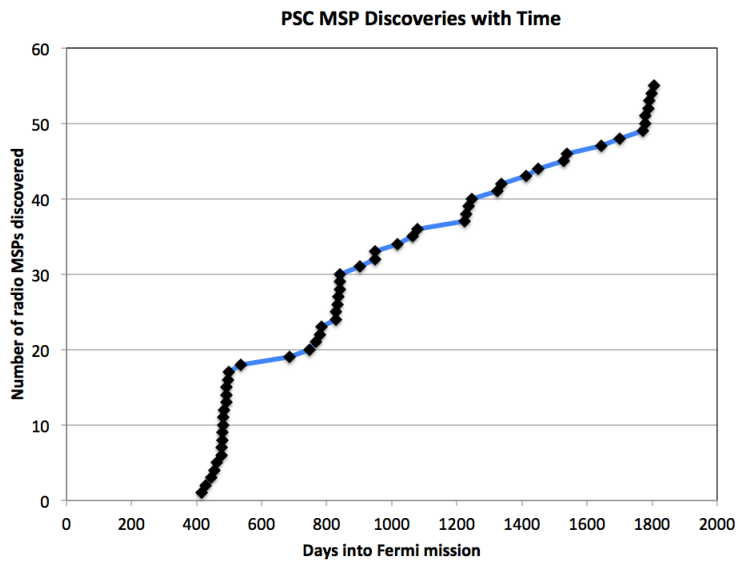

*Fig 3: Cumulative view of the MSPs discovered in radio, searching within promising unidentified* Fermi *sources. Once the potential for discovery of such a technique was realized, the MSP growth has been continuous and no flattening is in sight. (Courtesy of Elizabeth Ferrara)*

More searches are being conducted focusing on the unidentified sources with high "*pulsarness*" listed in the second *Fermi* source catalog ( Nolan et al., 2012, hereafter referred to as 2FGL) and the rate of discovery is impressive, as shown in figure 3.

While this is a major discovery on its own (since it increases significantly the number of known field MSPs), the sudden jump in the MSP number revived the interest of the radio-astronomical community in using these super stable clocks to detect nanoHertz gravitational waves (GW). The slight variations GWs may induce in the time of arrival of signals coming from widely separated pulsars will be correlated, thus allowing for the direct detection of GWs (e.g. Janet et al.,2005). Recent re-evaluation (Cordes and Shannon, 2012) of the method sensitivity show that a 5 y monitoring of (at least) 20 carefully selected MPSs stands a fair chance to detect the GW background and study its spectrum. Of course, a higher number of MSPs, ranging from 50 to 100 objects, would provide firmer results. Indeed, by finding so many new MSPs, *Fermi* increases the number of targets to be monitored, searching for GW induced variations.

The high success rate of *Fermi*'s treasure hunt also implies that MSPs could play a major role in accounting for unidentified Galactic gamma-ray sources, especially the faint, high-Galactic latitude sources.

Normal young pulsars, however, should not be totally forgotten. Camilo et al. (2012) for instance, during the radio coverage of 1FGL J2030.0+3641, found a middle-aged radio pulsar that was immediately detected also in gamma rays . Indeed, as stated by Camilo et al. (2012), the fact that so few young isolated pulsars have been found in LAT sources (which are mostly along the plane) is a testament to how good the radio surveys were.

Meanwhile, blind searches were conducted in newly discovered, low-latitude pulsar-like LAT sources, selected on the basis of their *"pulsarness"*. Eight radio-quiet INSs were found quickly (Saz Parkinson et al., 2010) and two more were later added (Saz Parkison et al., 2011)

Although optimized through the time-differencing techniques, blind searches on ever fainter LAT sources had to cover progressively longer time spans becoming very computer intensive, thus hampering the *Fermi*-LAT's potential for discovery. To overcome the limitation in computing power, the Albert Einstein Institute in Hannover brought their supercomputer (mainly devoted to the search for gravitational waves) into play, coupled with a new hierarchical search method originally aimed at detecting continuous gravitational waves from rapidly rotating neutron stars. The reward was immediate. Nine INSs were quickly found (Pletsch et al., 2012a). PSR J1838-0537 was added later (Pletsch et al., 2012b) and its timing analysis shows that in September 2009 the pulsar suffered the largest glitch seen so far in any gamma-ray-only pulsar.

Also, standard analysis using radio pulsar timing parameters continued, relentlessly following each potentially interesting pulsar since, if a pulsed signal is present, its significance grows with time and eventually reaches the 5-sigma level needed to announce a detection.

In view of the results obtained, the selection criteria used to build the list of pulsars under radio monitoring were revised, lowering the threshold $\dot{E}$ with a proportional increase in the number of pulsars under continuous investigation.

A lot of work was also devoted to improving the gamma-ray analysis technique. A weighting algorithm was introduced to assign to each photon a weight according to its (energy-dependent ) probability to come from a given pulsar (Kerr, 2011). Such a weighting algorithm reduces the trials previously needed to optimize the extraction region together with the energy range and results in an enhanced sensitivity to the pulsation detection.

In parallel, Ray et al. (2011) improved the analysis techniques by applying a maximum likelihood method to extract pulse times of arrival (TOA) from unbinned photon data. Using new phase-connected pulse timing solutions for the first group of 16 gamma-ray selected pulsars, they were able to improve the source positioning, rivaling the X-ray in localization accuracy. Moreover, continuous folding unveiled the presence of glitches for PSRs J0007+7303, J1124-5916 and J1813-1246 (Ray et al., 2011) joining several other radio pulsars for which gamma-ray folding had already highlighted glitches, starting with the very first detection of a glitch from PSR B1706-44 in just 10 weeks of data (Saz Parkinson, 2009). Indeed, it turns out that continuous gamma-ray coverage, coupled with the sensitivity of folding to pulsar parameters, is a powerful way to unveil pulsar glitches.

## 4. Towards the *Fermi* gamma-ray pulsar revolution

After so many advancements, it was time for a second pulsar catalog, based on 3 years of *Fermi* data. The number of pulsars detected was already beyond the most optimistic guesses published prior to launch. While Thompson (2001) bracketed the expectations between 30 and 100 objects, when the *Fermi* three-year observation database was frozen, 117 pulsars met the 5-sigma pulsation significance threshold (Abdo et al., 2013; hereafter 2PC). Figure 4 shows the pulsar positions against the 5-year *Fermi* image of the gamma-ray sky. Of these, 42 are radio-loud pulsars, 35 are radio-quiet, while 40 are MSPs, 20 of which were found through radio searches within unassociated *Fermi* LAT sources. Indeed, by the time the catalog was ready for publication, the number of MSPs discovered within unidentified sources had grown to 46 (Ray et al., 2012), 34 of which have been seen to pulsate in gamma rays .

Regarding the radio-quiet pulsars, a little clarification is in order. Out of the 36 INSs discovered by *Fermi* through blind searches, two (PSR J1741–2054 and PSR J2032+4127) were subsequently detected in radio (Camilo et al., 2009) and are thus counted as radio pulsars. On the other hand, PSR J1907+0602 (Abdo et al., 2010h) and PSR J0106+4855 (Pletsch et al., 2012d), both of which have been detected with exceedingly low radio fluxes, remain in the radio-quiet (or radio-faint) class. This class, then, contains 34 *Fermi* pulsars to which Geminga (as a prototype example of a truly radio-quiet pulsar) should be added.

The 2PC represents a milestone in pulsar astronomy. First of all, it establishes *Fermi*-LAT as a powerful pulsar discoverer. Indeed, half of the pulsars listed in the 2PC where not known prior to the launch of Fermi and they have been either discovered by *Fermi* through blind searches, or thanks to *Fermi* by radio targeted searches of *Fermi* LAT unassociated sources . Moreover, the 117 entries are divided almost exactly "in partes tres" between young radio-loud pulsars, young radio-quiet pulsars and MSPs. Indeed, the most dramatic advancements have been achieved in the field of MSPs. While, before *Fermi*, 70 MSPs were known outside globular clusters, now there are 120, 39 of which are part of the 2PC. The 40$^{th}$ *Fermi* MSP is J1823-3121A, located within the globular cluster NGC6624, which represents the first detection of gamma-ray pulsations from a MSP in a globular cluster (Freire et al., 2011).

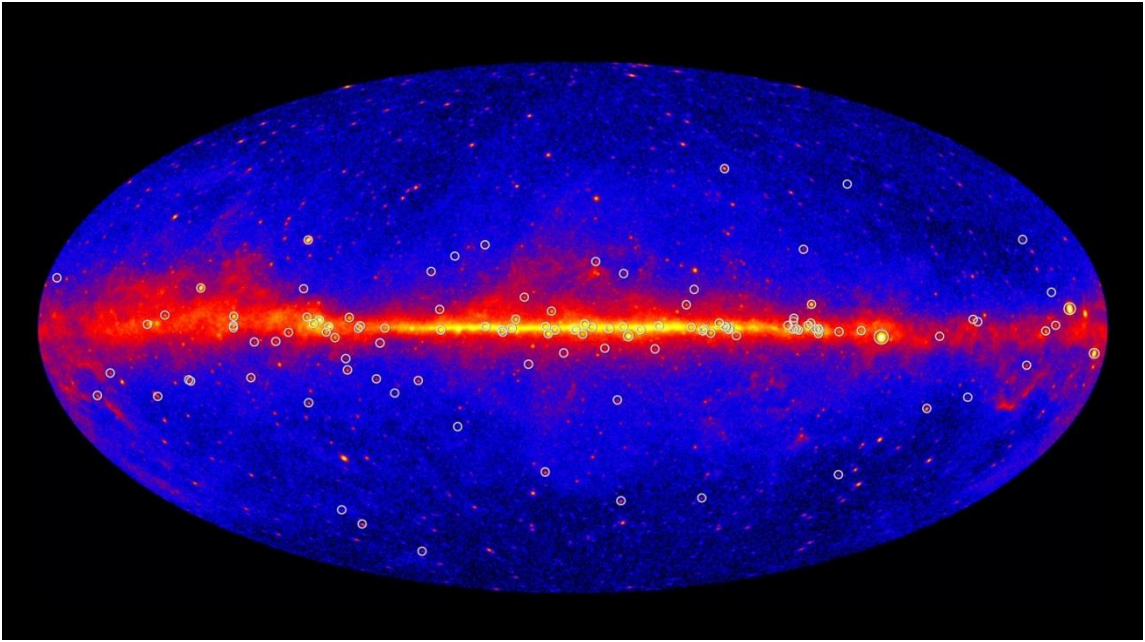

Figure 4: *Fermi 5-year sky map showing the positions of the 117 pulsars listed in the 2PC. Image from NASA/DOE/Fermi LAT Collaboration* http://svs.gsfc.nasa.gov/vis/a010000/a011300/a011342/

MSPs found in Fermi error-boxes are a shorter-period, more energetic population than radio-selected ones (Ray et al. 2012). Indeed gamma-ray MSPs dominate for P< 0.003 s. In addition, Roberts (2013) remarks that MSPs found in *Fermi* sources have dramatically increased the number of "Black Widow"-type systems where the MSP is driving the evaporation of its tightly-bound very light binary companion.

Thus, the unexpected split into three pulsar classes makes 2PC noteworthy both in quantity and diversity. This is highlighted in tables 1,2, and 3 which summarize the parameters of the 117 2PC pulsars divided into radio-loud, radio quiet and MSPs.

When looking at the 2PC numbers, one must also consider the sensitivity issue since different search techniques, as well as different sky locations, do result in different sensitivities. As shown by Dormody et al. (2011), blind searches are about 2.5 times less sensitive than folding gamma-rays with a known ephemeris. Thus, the similar number of detections among radio-loud and radio-quiet pulsars tells us that the parent population must contain more radio-quiet pulsars, as expected based on the OG model. (e.g.

Romani and Yadigaroglu 1995 foresaw that the radio-quiet gamma-ray emitting INSs should account for half of the young NSs, but that only 19% should be visible both in gamma-ray and radio wavelengths).

**4.1 The Galactic distribution of pulsars revealed by Fermi**

Although hard to distinguish from their gamma-ray light curves and spectra, the three pulsar families have markedly different Galactic distributions, and different average fluxes.

The Galactic distribution of the different classes of 2PC pulsars, given in figure 5, clearly shows that young pulsars (both radio-loud and radio-quiet) are clustered around the Galactic plane, where they are much harder to detect due to higher Galactic background radiation, while MSPs (be they gamma-ray pulsars or new radio pulsars discovered in *Fermi* error boxes) are distributed all over the sky, pointing to a relatively local origin. However, the paucity of MSPs close to the Galactic plane is probably an observational bias since the *Fermi* unidentified source radio programs preferentially select sources at high Galactic latitude, avoiding regions where gamma-ray diffuse emission is higher and source confusion more likely.

The similarities/differences between the three pulsar classes can be immediately gauged by plotting their LogN-LogS distributions, (see Figure 6). While young pulsars (both radio-loud and radio-quiet) have comparable fluxes and a number-flux slope in the range -0.7 to -0.9, the MSPs number-flux distribution is definitely steeper (slope = -1.6) with a much lower average source flux. According to a straightforward geometric interpretation of the LogN-LogS plot, a Galactic cylindrical distribution with no boundaries would yield a -1 slope, while an homogeneously filled sphere would be characterized by a -1.5 slope. Thus, young pulsars seem to trace a Galactic disk-like population that lacks faint sources, probably due to the high Galactic background, while MSPs point to a spherical distribution of intrinsically fainter sources in our Galactic neighborhood.

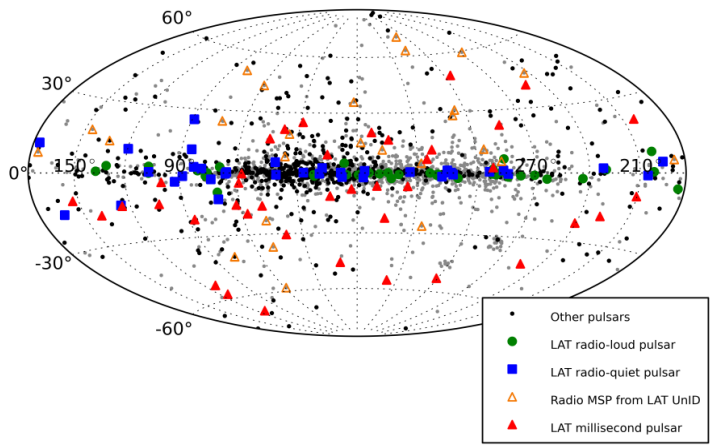

Figure 5: The *Galactic distribution of 117* Fermi *pulsars: radio-loud pulsars in green, radio-quiet in blue, MSPs in red. Empty triangles represent radio MSPs discovered within* Fermi *unidentified sources while filled triangles are MSPs detected as gamma-ray pulsars. Black dots indicate 710 radio pulsars that were (unsuccessfully) phase folded with radio ephemerides provided by the "Pulsar Timing Consortium". Grey dots indicate 1337 pulsars outside globular clusters for which phase folding was not performed. (from 2PC)*

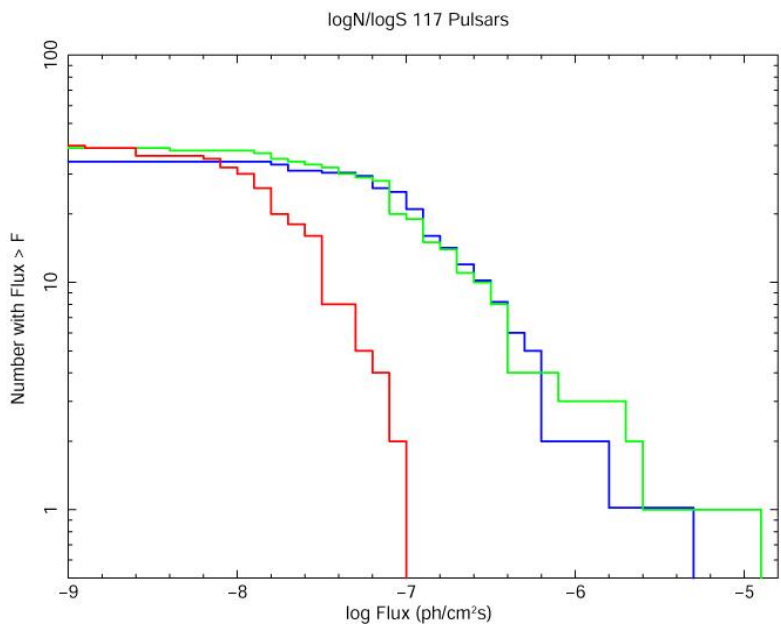

*Figure 6: LogN-LogS plot for the three classes of gamma-ray pulsars: green line indicates radio-loud, blue line radio-quiet, red line MSPs*

**4.2 General population properties: are radio-loud pulsars different from radio-quiet ones ?**

To put in context the 117 *Fermi* pulsars, it is useful to plot them on the pulsar $P$-$\dot{P}$ diagram. This is given in Figure 7, extracted from the 2PC. At first glance it is easy to note that all the detections lie above the line corresponding to $\dot{E}=10^{33}$ erg/sec, which could represent either an observational bias or a true death line for gamma-ray pulsar emission. Among the young gamma-ray pulsars a "segregation" effect is seen, with radio-quiet INSs dominating the $\dot{E}$ interval ranging from $10^{33}$ to $10^{35}$ erg/sec while the opposite is true for $\dot{E}$ > $10^{37}$ erg/sec, where only 1 of the nine pulsars detected by *Fermi* is radio-quiet. This feature, first noted by Ravi et al. (2010), can be linked to the radio and gamma-ray beaming factors. For high-$\dot{E}$ pulsars the beams should be similar both in sky coverage and in location, and be rather high in the pulsar magnetosphere, while for low-$\dot{E}$ pulsars the radio-emitting region should migrate towards the NS surface, thus shrinking the sky coverage of the radio beam and resulting into a higher percentage of radio-quiet INSs. Such a beaming evolution is also discussed by Watters and Romani (2011) in their population synthesis simulation.

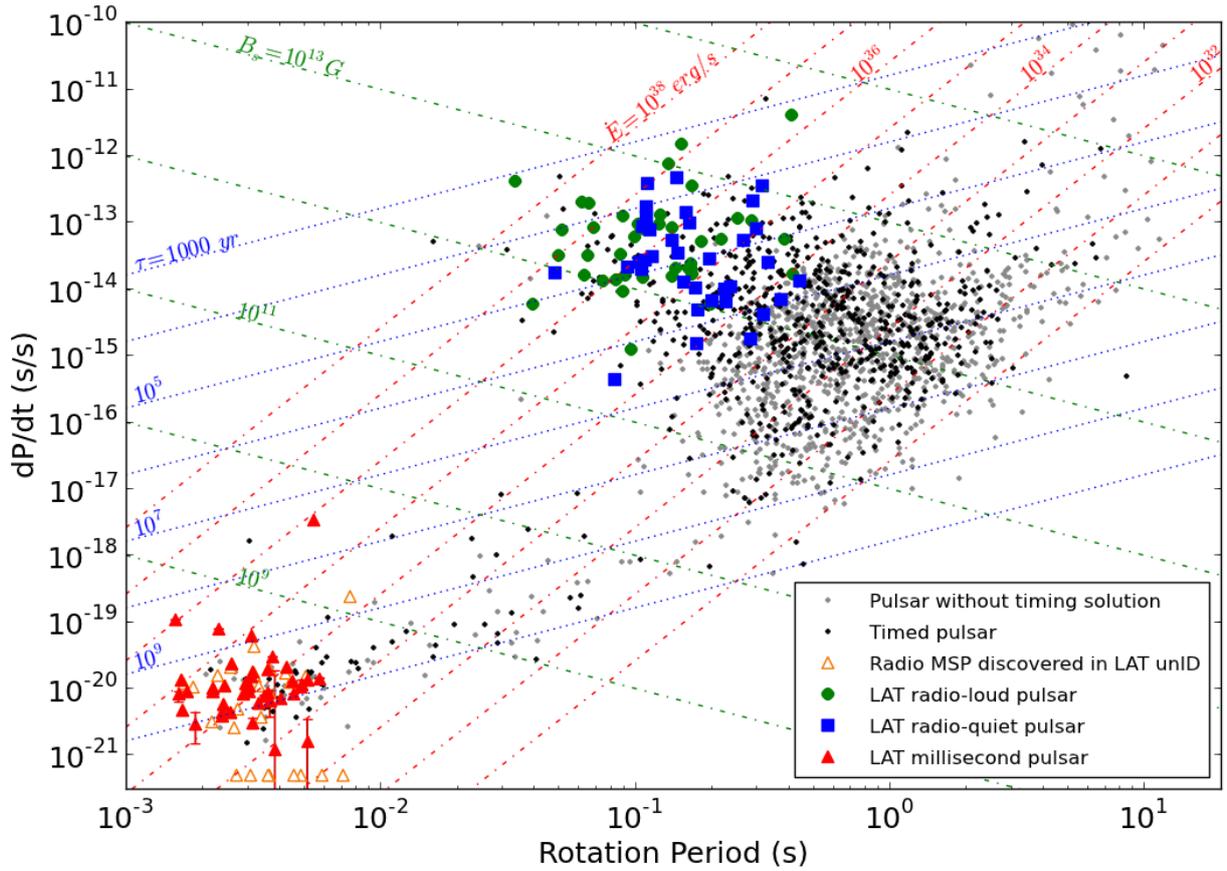

*Figure 7: P-Ṗ distribution of the 117 Fermi pulsars (symbols as for figure 5) plotted together with the entire radio pulsar sample known today (710 black dots represent timed pulsars which were phase folded, but not detected, while 1337 grey dots represent pulsars without timing solution). Lines of constant rotational energy loss ($\dot{E}_{rot} = -4\pi^2 I \dot{P} P^{-3}$), characteristic age ($\tau_c = \frac{P}{2\dot{P}}$) and surface B field*

*($B_s = (1.5I\ c^3 P\dot{P})^{1/2}/2\pi R_{NS}^3$) are also shown. Recently discovered MSPs, for which no Ṗ has been measured, are plotted at Ṗ = 5 x 10²². All the Fermi-LAT pulsars lie above $\dot{E}$ = 10³³ erg/sec (from 2PC)*

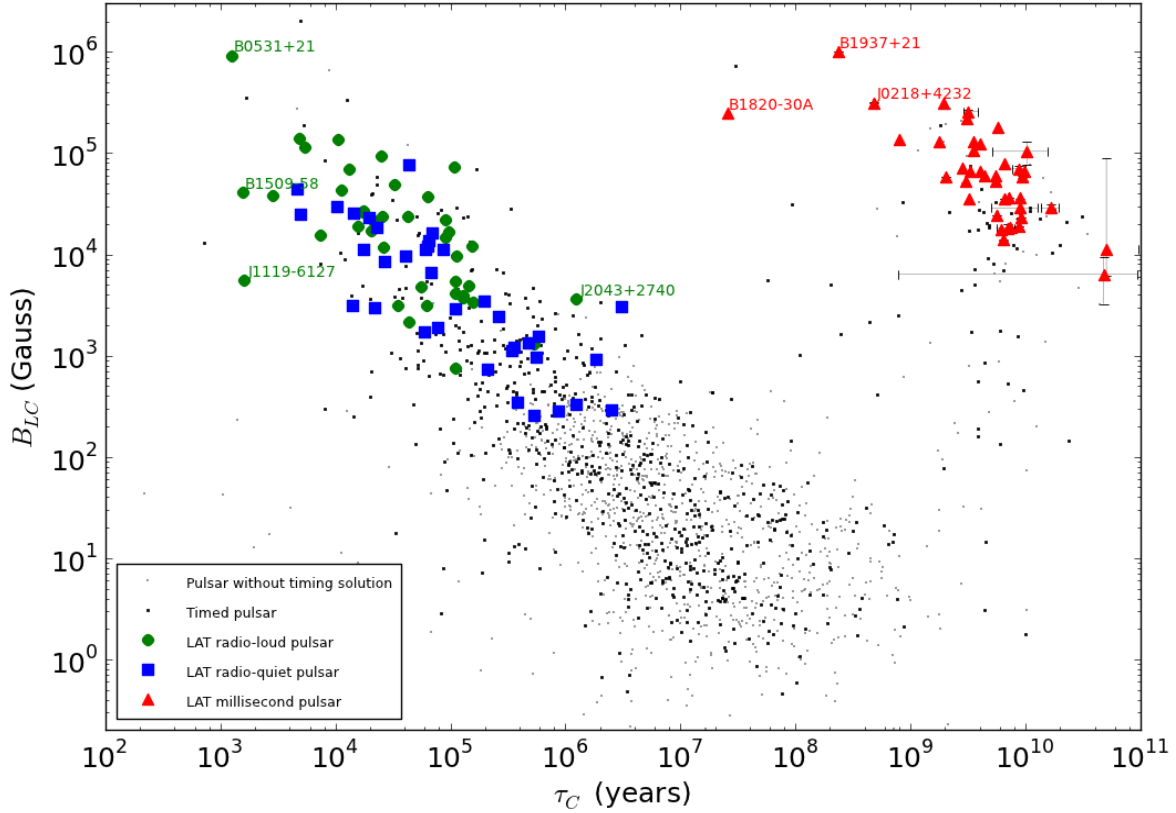

Figure 8: The magnetic field at the light cylinder $B_{LC} = 4\pi^2(1.5I\dot{P})^{1/2}(c^3P^5)^{-1/2}$ is plotted against the INS characteristic age $\tau_c = \frac{P}{2\dot{P}}$ . Radio-loud and radio-quiet pulsars are distributed between $10^2$ and $10^6$ G while MSPs are more clustered in the $10^4$-$10^6$ G region (from 2PC, symbols as for figure 5)

To understand the *Fermi* pulsars one of the key parameters seems to be the magnetic field at the light cylinder. As we have already remarked, while the surface B-field of young pulsars is $10^4$ higher than that inferred for old, recycled ones, the different sizes of their corotating magnetospheres result in similar B-field at the light cylinder. This is shown in Figure 8 where the value of the magnetic field at the light cylinder is plotted against the characteristic age of the pulsars. Once more, the radio-loud /radio-quiet segregation is apparent with the majority of the middle-aged "young" pulsars being radio-quiet, while the opposite is true for very young pulsars. Such an effect could result from an observational bias hampering the detection of young, but far-away and thus faint, radio-quiet, very young pulsars. On the other hand, if true, such behavior could also be interpreted in the framework of the migration of the radio beaming described above, whereby older and less energetic pulsars are more likely to be spotted as gamma-ray emitters, rather than radio ones.

A similar segregation is also seen in the young pulsar spectral parameters. All 117 2PC pulsars have been fitted in a consistent way using a power law with exponential cutoff spectrum. For each pulsar, a power law index, as well as a value for the cutoff energy, has been computed. While the power law indices inferred for MSPs and young pulsars do show positive correlation with their rotational energy loss, when one plots the cutoff energy as a function of the magnetic field at the light cylinder (as in figure 9), radio-quiet pulsars dominate in the region characterized by low cutoff and low magnetic field.

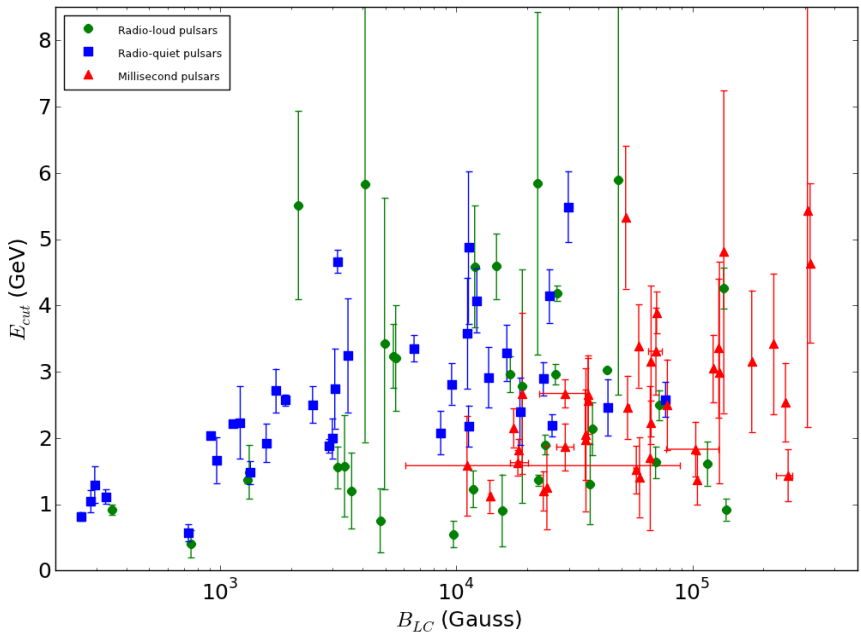

*Figure 9: Cutoff energy as a function of the magnetic field at the light cylinder (symbols as in figure 5)*

A further effort to characterize radio-loud versus radio-quiet-INSs has been done through their X-ray emission. A thorough analysis of all the X-ray data available for *Fermi* pulsars has been performed by Marelli, 2010, Marelli et al 2011, later updated within the 2PC) to search for correlation (if any) between the distance independent ratio $F_\gamma/F_x$ (gamma-ray flux over X-ray flux) and the pulsar age. As there is significant spread in the $F_\gamma/F_x$ values that spans 3-4 orders of magnitude for objects of similar age and energetics, no correlation was found, although inspection of the X-ray flux data revealed that radio-quiet pulsars tend to be underluminous in X rays.

**4.3 The efficiencies of gamma-ray pulsars : Aged to perfection**

Plotting the gamma-ray luminosity as a function of rotational energy loss of pulsars does not yield a clear picture. Although the plot (shown in Figure 10) is hampered by the lack of distance estimates for the majority of radio-quiet neutron stars , a different trend is apparent for the young pulsars as opposed to the old, recycled ones. While young pulsars seems to follow a $\sqrt{\dot{E}}$ trend, MSPs prefer a steeper function, pointing to a proportionality between $L_\gamma$ and $\dot{E}$. However, the presence of considerable scattering should not be overlooked. It is probably due to the combination of distance uncertainty with the assumption of a common $f_\Omega$ =1. The latter assumption is generally considered acceptable for an outer-magnetosphere fan-like beam(s) sweeping the entire sky. But it has been questioned by Pierbattista et al. (2012) who found a large spread of $f_\Omega$ among the different emission models and, in more general terms, between radio-loud and radio-quiet pulsars. Clearly, an $f_\Omega$ value optimized for each pulsar using the model yielding the best fit for its light curve, would be highly desirable.

However, when plotting the *Fermi* gamma-ray efficiency (i.e. the ratio between the gamma-ray luminosity and the pulsar rotational energy loss $L_\gamma / \dot{E}$) as a function of the pulsar rotational energy loss as in Figure 11, a clear trend appears showing that old pulsars, which can count on a much smaller energy reservoir, are more efficient in converting their rotational energy loss into gamma rays .

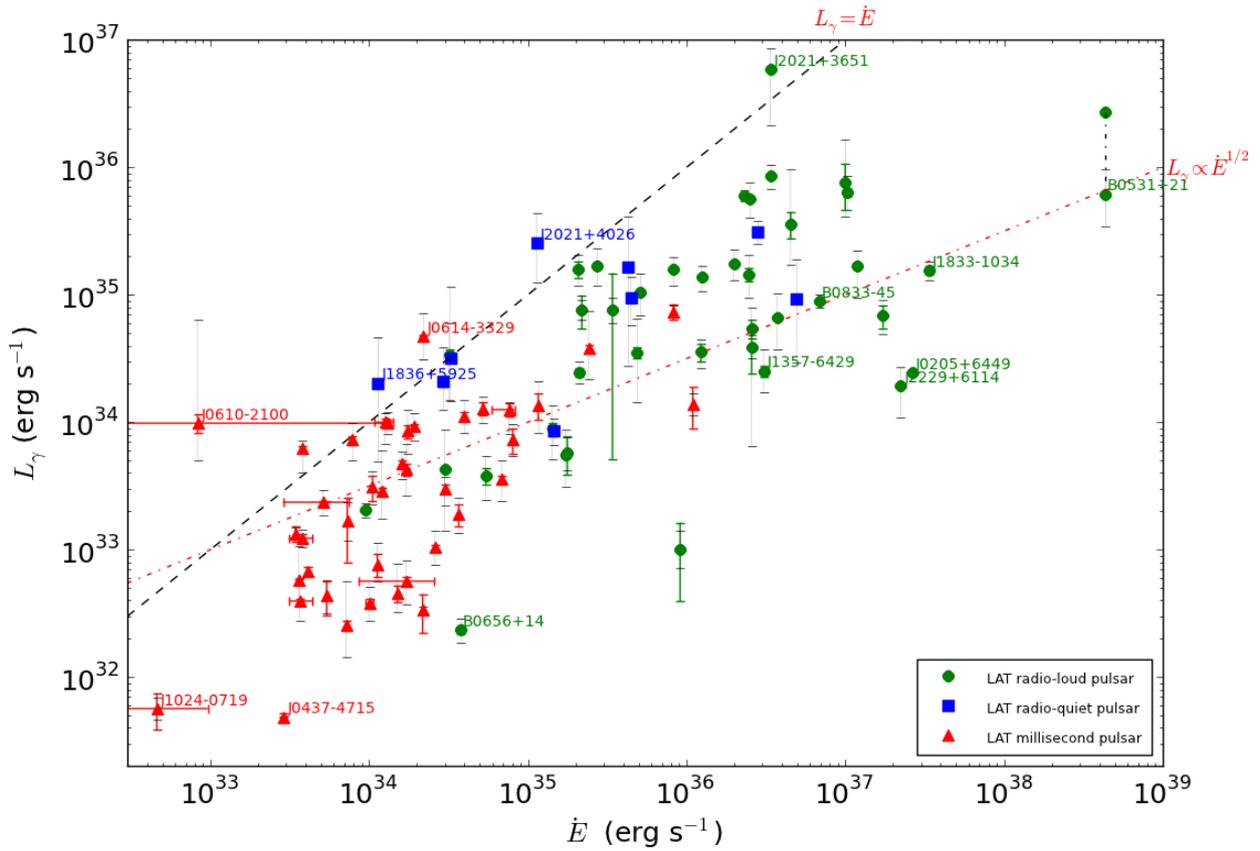

*Figure10: Gamma-ray luminosity ($L_\gamma = 4\pi d^2 F_\gamma f_\Omega$, computed assuming $f_\Omega$=1) as a function of the rotational energy loss. The difference in slope between the red cloud of MSPs and the green cloud of radio pulsars is evident. The paucity of radio-quiet pulsars is due to the lack of distance information. A dashed line represents $L_\gamma = \dot{E}$, a rather extreme case used to highlight pulsars whose distance must be significantly overestimated or which require a much narrower beaming factor. A dash-dotted lines follows $\sqrt{\dot{E}}$ . The Crab pulsar, PSR B 0531+21, is the only pulsar detected as a very bright X-ray source. The lower point represents the gamma-ray luminosity of the Crab pulsar, while the upper one indicates the total luminosity including X-rays (from 2PC, symbols as for Figure 5)*

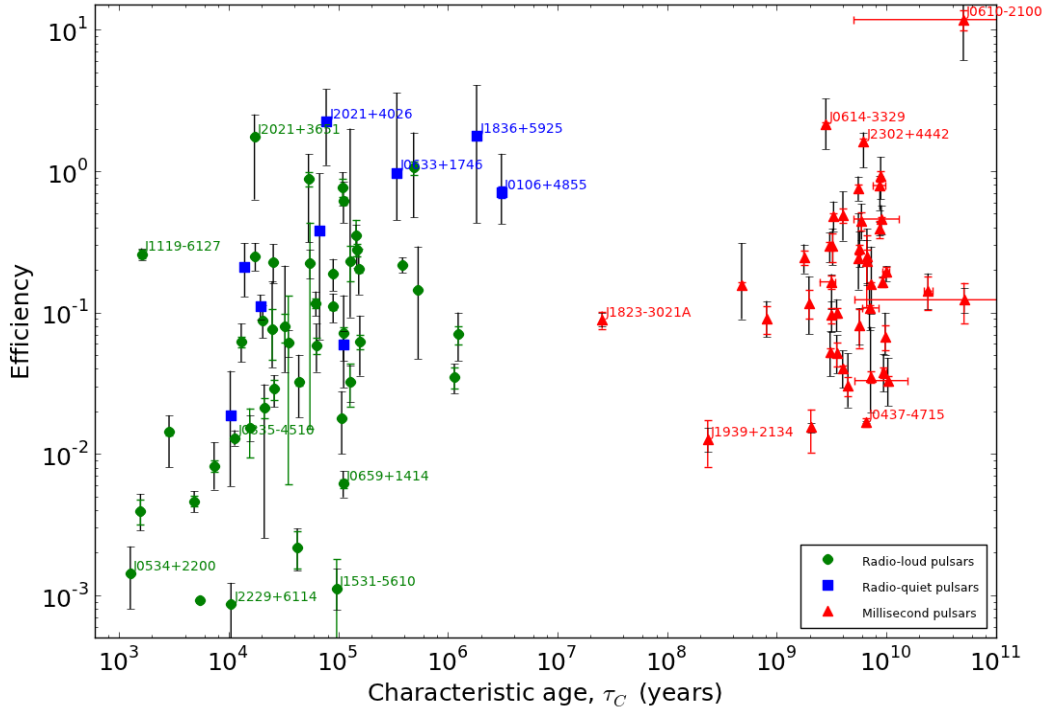

*Figure 11: Gamma-ray efficiency ($L_\gamma/\dot{E}$) plotted as a function of pulsar's characteristic age, showing that older and less energetic pulsars are more efficient in converting their rotational energy loss into high-energy gamma rays (symbols as in figure 5). The number of pulsars with an efficiency near or above 1 is certainly due to overestimated distances as well as to a too-large beaming factor (here assumed to be 1 for all pulsars). Such a hidden beaming uncertainty is certainly responsible for the unrealistically high efficiency computed for Geminga (J0633+1746), for which the distance is well known. Courtesy of David Smith, using data from 2PC.*

**4.4 Do gamma-quiet pulsars exist?**

So far we discussed the characteristic of the INSs detected by *Fermi*. What about those not seen? The point is a non-trivial one since, given the beamed nature of the pulsar emission, it is natural to expect a geometric configuration such that radio emission is detected but gamma-ray pulsation is not. According to Romani and Yadigaroglu (1995), for instance, 8% of the pulsar population should be gamma-ray quiet.

Considering the case of MSPs, Guillemot and Tauris (2013) find that the most energetic MSPs may go undetected in gamma-rays owing to unfavorable geometry.

Are there pulsars highly ranked on the basis of $\frac{\sqrt{\dot{E}}}{4\pi d^2}$ that are not detected by *Fermi*? 28 of the 64 known radio pulsars with $\dot{E} > 10^{36}$ erg/sec have not been detected, and, comparing their upper limits with the "expected" flux (on the basis of their ranking) a number of interesting gamma-quiet candidates appear to be present. However, as discussed in 2PC as well as in Romani et al. (2011), distance uncertainties, confusion with nearby sources and the possible presence of significant timing noise make it difficult to provide a clear-cut answer. A few INSs have expected gamma-ray flux values well above the current upper limits but none is yet a fully convincing gamma-quiet radio pulsar.

**4.5 Light Curves and their interpretation**

Upon inspecting the light curves of the 117 gamma-ray pulsars, one immediately realizes that the majority of the pulsars (70% of the young pulsars and 60% of the MSPs) have two peaks (respectively P1 and P2, where P1 is defined as the one soonest in phase after the radio main peak), and that the ratio P2/P1 increases with energy, indicating a harder second peak. Many double-peaked pulsars display a crescent-type light curve with significant emission between the two peaks. For Crab and Vela a third peak emerges from the interpeak bridge emission. However, the third peak of Vela is seen to move in phase as a function of energy (Abdo et al., 2010e), an effect not expected in the current geometric models.

For the radio pulsars, the lag between the radio signal and P1 is carefully evaluated since it is one of the important parameters in characterizing a pulsar's multi-wavelength behavior. Indeed, the lag between the radio peak and P1 is greater for MSPs than for young INSs, indicating that a smaller magnetosphere implies a stronger aberration of the radio pulses.

Such a wealth of information on pulsar light curves represents a new challenge for theoreticians who try to constrain the geometry of pulsars as well as the relevant magnetospheric physics. Starting from the location(s) of the emitting region(s), namely polar cap (PC), outer gap (OG) and slot gap (SG), or its variation two-pole caustic (TPC), in a dipole geometry, one can build an "atlas" of predicted gamma-ray light curves to be compared to observations. The computed light curves are sensitive to both the magnetic axis inclination ($\alpha$) and the viewing angle ($\zeta$), and with no a priori knowledge on such variables, all the combinations should be considered. Figure 12 provides an example of emission patter phase plots computed for a given magnetic inclination ($\alpha$= 45°) for all possible viewing angles. The actual pulsar light curve, obtained cutting the phase plot for a definite value of $\zeta$, can be used to select the best fitting value of $\zeta$ on the basis of the different models. Conversely, external inputs, such as radio polarization angle or X-ray morphological study, can help to constrain the parameter space. The first gamma-ray atlas was compiled by Watters et al. (2009) for vacuum dipole field geometries. Bai and Spitkovsky (2010) considered numerically-modeled 'force-free' geometries, while Venter et al (2009) concentrated on the newly-established MSPs. In general, OG models yield better fits, but they are not able to account for all the detected pulsars. Lower-altitude emission is preferred for a sizable minority of pulsars, especially those MSPs with aligned radio and gamma-ray light curves. In parallel, emission beyond the light cylinder continues to be a viable alternative (as discussed by Petri, 2011).

Pierbattista (2010) extended the atlas approach to also include radio light curves by trying to constrain the $\alpha$ and $\zeta$ parameters on the basis of a joint radio-gamma fit. Although promising, the procedure still needs fine-tuning in order to avoid being driven by the radio data. Once a comprehensive gamma-ray and radio light curve atlas is in hand, it can be used to synthesize a pulsar population to compare with the *Fermi* findings (Pierbattista et al., 2012).

Not surprisingly, the OG model, with its extended beaming, can easily account for the observed number of *Fermi* detections while the narrow PC beams can only account for a few pulsars, and the SG need a boost in efficiency to account for the observed set of LAT detections. The population synthesis, however, fails to reproduce the LAT results for high $\dot{E}$ pulsars. All the models predict too few high $\dot{E}$ pulsars and cannot explain the high probability of detecting energetic radio-loud pulsars.

In a nutshell, none of the models proposed so far is able to account for the phenomenology of the observed *Fermi*-LAT pulsars: OG/SG are generally better but not adequate to fit all objects. Some pulsars can be reasonably well fitted by more than one model, some by none.

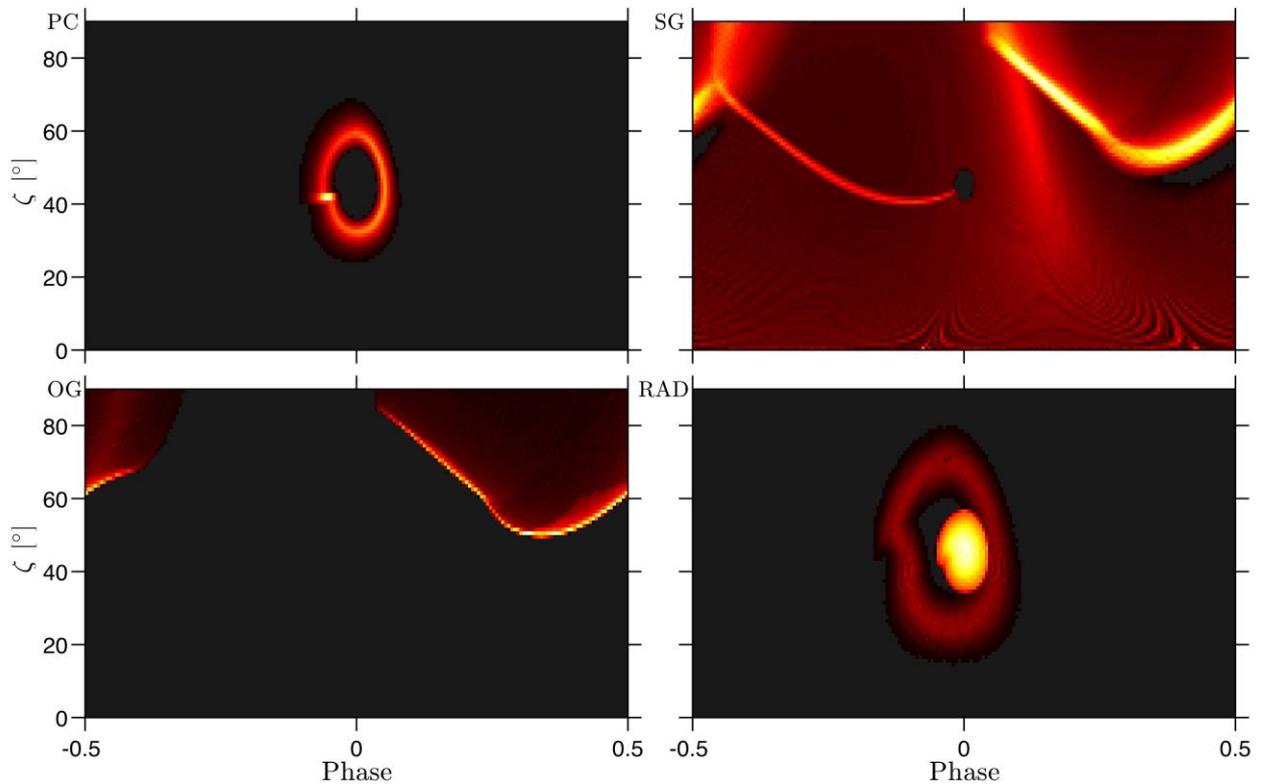

Figure 12 *Emission patter phase-plots computed, respectively, for PC, SG, OG, and radio core plus cone models. For PC and radio models the phase-plots have been obtained for a magnetic field B=10$^{12}$ Gauss and spin period of 30 ms, while for SG and OG models the phase-plots were computed for gap widths of 0.04 and 0.01 respectively. All plots have been obtained for a magnetic obliquity α = 45◦. The emission intensity decreases from yellow to black. (Pierbattista et al, 2012)*

## 5. Paradigm Lost: winning the pulsar revolution

While the 2PC was being written, the gamma-ray pulsar family gained 15 more members that are listed in Table 4 of the catalog. 11 of the new entries are MSPs while 4 are young radio pulsars. Altogether, as of September 2013, 132 NSs have been detected in gamma rays . Their parameters can be found in
https://confluence.slac.stanford.edu/display/GLAMCOG/Public+List+of+LAT-Detected+Gamma-Ray+Pulsars

Thanks to the extremely successful synergy between radio and gamma-ray astronomy to find, characterize and phase-fold MSPs, the number of MSP discoveries currently accounts for two thirds of the new pulsars. However, one such MSP was not found through radio searches but rather through an original approach exploiting at once gamma-ray and optical astronomy. Performing blind searches for MSPs in binary systems is vastly more difficult than searching for young isolated pulsars, since the search must cover a much larger frequency interval, scanning for (at least three) additional unknown orbital parameters. This becomes an impossible task, even for the most powerful supercomputer. The source 0FGL J1311.9-3419, bright enough to be already listed in the very first *Fermi* Bright Source Catalog (Abdo et al., 2009i), however, yielded a valuable hint through its optical emission. In an observing campaign aimed at finding optical variation of MSPs binary companions within unidentified Fermi sources, Romani (2012) detected a quasi-sinusoidal 93-minute modulation that was thought to arise from a "black widow" system (where the MSP irradiates its low mass companion leading, eventually, to its evaporation). Indeed, the 93-m period is the shortest known for such a binary system. By focusing on the coordinates of the variable optical source and exploiting the tight constraints placed on the orbital parameters, Pletsch et al. (2012c) were able to detect a 2.5 ms MSP, the first found through a gamma-ray blind search, an historic accomplishment. However, PSR J1311-3430 is not the long sought radio-quiet MSP: radio astronomers went back to the source, which had been already observed to no avail and, with considerable effort, Ray et al. (2013) found the radio signal which is present only during a small fraction of the radio observations. Further optical studies allowed Romani et al. (2012) to constrain the mass of the NS to be $> 2.1\ M_\odot$.

A similar optical strategy was applied to 0FGL J2339.8-0530. Variability pointed to a tentative orbital period (Romani and Shaw, 2011) that was the driver behind a successful radio search, soon leading to a gamma-ray detection, listed in the 2PC. Those success stories rest on new multi-wavelength strategies that will certainly bring more results in the future.

**5.1 2013: enters Citizen Science**

Blind searches for gamma-ray pulsars are computationally intensive, thus any additional computer resource is most welcome. To increase (at no cost) the computer power to be used in their searches, Pletsch et al. (2013) exploited the potential of the volunteer distributed computer system Einstein@Home (Allen et al, 2013), a well-known Citizen Science project that was started in 2005 to search for gravitational waves in data collected by the LIGO-Virgo Collaboration. By downloading the Einstein@Home software, volunteers agree to devote the power of their computer, when not in use, to perform the Einstein@Home tasks. With more than 300,000 volunteers, Einstein@Home is one of the most popular Citizen Science projects and averages a total computing power of 1 PFLOPS, comparable to the largest supercomputers. Since 2009, the system has been adapted to analyze radio telescope data, yielding several pulsar discoveries (Allen et al, 2013), and is now being used to perform blind searches on *Fermi* data. So far, four young pulsars have been discovered in unidentified 2FGL sources. Their parameters are characteristic of energetic objects relatively nearby, but none of them has yet been detected in radio. Considering the shortage of energetic radio-quiet pulsars, this result is remarkable. Two of the new pulsars have already been seen to glitch: PSR J1522-5734 glitched once, while PSR J1422-6138 glitched twice, proving that the system works well in spite of those timing jumps. The additional computer power provided by E@H combined with an improved search technique is a good omen for more findings, hopefully also of the long-sought radio-quiet MSP.

**5.2 Breaking a decade-long paradigm**

The discovery of the Crab Nebula variability has been one the most astonishing results in high-energy gamma-ray astronomy. On September 2010, a sudden enhancement of the overall Crab flux was reported first by AGILE and immediately confirmed by *Fermi* (Tavani et al., 2011, Abdo et al., 2011). A quick sequence of radio, X-ray, and optical observations made it clear that the pulsar was behaving normally. Indeed, in gamma rays the pulsed flux from the Crab was also unchanged, leaving the nebula as the only suspect for the flux increase. Moreover, the short time scale of the variability pointed to a quite small region of interest, possibly next to the pulsar. To achieve the highest angular resolution to study the interior of the Crab Nebula, next to the pulsar, the *Chandra* X-ray observatory and the Hubble Space Telescope (HST) were immediately triggered for Target of Opportunity (ToO) observations which were performed a few days after the event. The high-resolution optical and X-ray images failed to show any dramatic change in the notoriously active Crab nebula inner region. A search in both AGILE and *Fermi* data did prove that enhancements had been detected previously, making it clear that the Crab's Sept. 2010 flaring episode was not a unique event but rather a recurring one. A massive ToO campaign was organized, waiting for the next flares which were recorded in April 2011 (Striani et al., 2011, Bueher et al., 2012, Weisskopf et al., 2013) and in March 2013 (Mayer et al., 2013) when both Chandra and HST repeatedly observed the nebula. Once again, comparing images taken before, during and after the flare, nothing obvious was seen to change. The Crab flares shine only in gamma rays , and such events are possibly linked to sudden particle acceleration, driven perhaps by magnetic reconnection. Hopefully the coming years will provide more flares to test the various proposed theories.

In spite of the "*pulsarness*" definition we have used so far, variability in INSs and in their surroundings may not be all that exceptional. Allafort et al. (2013) recently reported the detection of a significant flux variation, this time a decrease, from PSR J2021+4026, a radio-quiet INS in the Cygnus region with an X-ray counterpart (Weisskpoft et al., 2011). Judging from its light curve, PSR J2021+4026 is similar to Geminga and recently joined Geminga and CTA-1 in the very small club of radio-quiet gamma-ray pulsars seen to pulsate in X rays (Lin et al., 2013). The gamma-ray flux decrease, which took place in less than a week, is associated with a 4% increase in the pulsar's spin-down rate and a change in the light curve . The timing parameters of the pulsar have changed in a way never before seen in gamma rays . A jump in $\dot{P}$ is typically followed by a recovery pattern. Moreover, flux variability has never been associated with a glitch, despite repeated analyses during major Vela glitches to search for such behavior. Is the jump due to a shift in the magnetic field structure? If so, PSR J2021+4026 may have done it before, as AGILE reported variability from this gamma-ray source (Chen et al. 2011), although they could not link their finding to the pulsar. As puzzling as it may sound, these results may be heralding a new era where the pulsar steady flux paradigm is superseded.

## 6. Now what?

Figure 13 provides a snapshot of the current census of *Fermi* pulsars (divided between radio-loud, radio-quiet and MSPs) together with an overall view of 4 decades of pulsar studies in high-energy gamma rays.

While the growth during the last five years is dramatic, to say the least, it is apparent that the most recent detections have changed the family balance, making MSPs the dominant class among the gamma-ray pulsar types. This is the unexpected, surprising and really revolutionary result of the *Fermi* mission. Predicting the detection of more MSPs is easy on the basis of the discovery rate shown in Figure 3. The next pulsar catalog will be dominated by MSPs, many in black-widow systems.

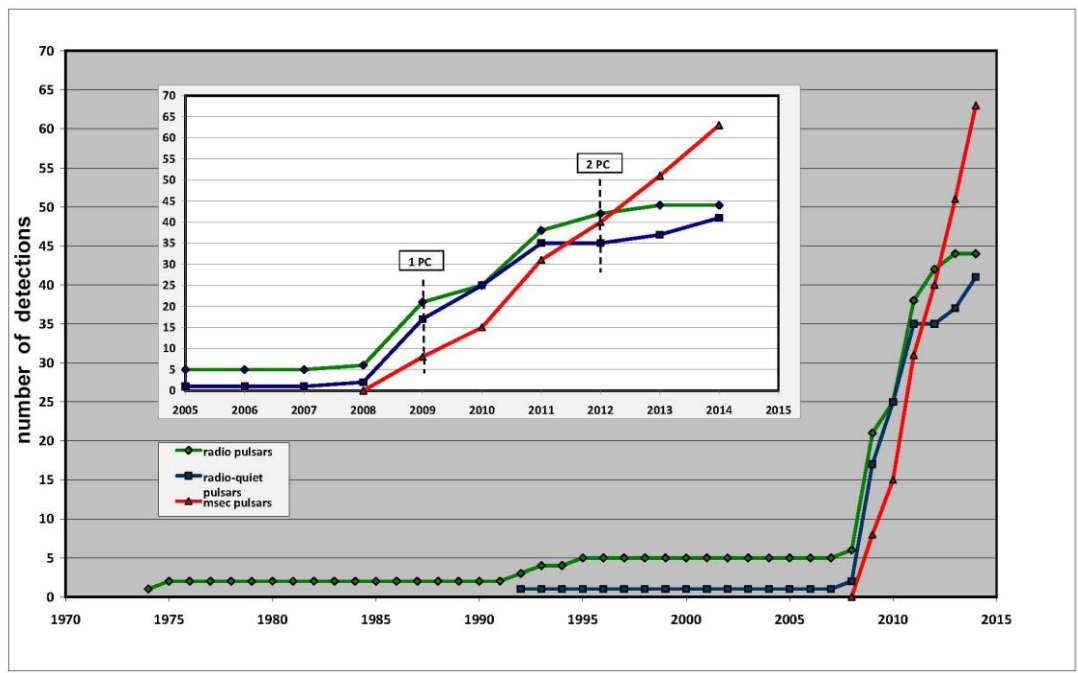

*Figure 13: Pulsar census for radio-loud pulsars (green), radio-quiet pulsars (bleu), and MSPs (red). The inset zooms in on the dramatic advancements of the last five years. The pulsar accounting published in the first (1PC) and second (2PC) pulsar catalogs are highlighted.*

Together with more MSPs, fainter young radio pulsars will join in (Hou et al., 2013) also owing to a major *Fermi* software update, which promises improved sensitivity at low energies (Atwood et al,2013). The statistical significance for faint sources builds up slowly, and endurance is required to spot the less energetic pulsars. However, this effort will be handsomely rewarded as the detections of even a handful of pulsars fainter than the current death line in the P-$\dot{P}$ diagram will have important implications in pulsar physics. Moreover, lowering the $\dot{E}$ threshold for gamma-ray detection would bring many more INS candidates into play, thus increasing their contribution to the overall Galactic emission, while also providing an obvious source for the overabundance of positrons detected by Pamela (Adriani et al., 2009), Fermi (Ackerman et al, 2012) and AMS (Aguilar et al, 2013).

The discovery of gamma-quiet INSs detected only at radio wavelengths is also an important goal since it will provide a missing piece of the puzzle for pulsar emission geometry. So far, all the INSs we know come in two flavors: radio-and-gamma and gamma-only. It is time to complete the picture with a few representatives of the gamma-quiet class. Detecting just unpulsed gamma-ray emission from a NS magnetosphere is also an intriguing possibility. But it will be hard to secure a convincing identification without a pulsar time signature.

By solving the riddle of dozens of previously unidentified gamma-ray sources, pulsars, be they young or recycled, prove to be the most promising candidates to account for a sizable fraction the remaining unassociated Galactic gamma-ray sources. By continuing to study unassociated sources, Fermi will certainly unveil many more surprising results.

Surprises may come also from the scores of INSs already detected in gamma-rays. While variability may be more common than previously thought, the combined use of many of the MSPs unveiled by *Fermi* may provide a tool to directly detect, at last, gravitational waves.

**Acknowledgements**

This paper is my own, mostly first-hand account of the 40 year-long struggle to detect and understand gamma-ray emission from neutron stars. During such a long time span I enjoyed working and discussing with many colleagues whose results and opinions I have tried to describe as accurately as possible.

I am indebted to the whole *Fermi* LAT collaboration who did a tremendous job in changing our view of the gamma-ray sky.  Working with the Galactic group has always been fun and stimulating. I have learned a lot from Alice Harding and Roger Romani, who have been the theoretical task force behind many Fermi papers and, quite rightly, have been awarded the 2014 Rossi prize for their contribution towards the understanding of gamma-ray pulsar emission. I am especially grateful to David Smith, who helped me to handle all the Fermi related-figures as well as the pulsar tables. I really appreciated his critical assessments and his clever suggestions. Elizabeth Ferrara and Pablo Saz Parkinson were kind enough to read and comment the manuscript. To them my most sincere thanks for their time and their patience.  Marco Pierbattista kindly  provided figures 2 and 12. I am also grateful to the Fermi Publication Board who granted permission to use unpublished data on the most recent pulsar census.

Table 1 Radio-Loud INSs

| JName | BName | Glat | Glon | P | Pdot | Edot | Lum | Dist | Age | Bs | B_LC | Comments |
|---|---|---|---|---|---|---|---|---|---|---|---|---|
| | | deg | deg | ms | s/s | erg/s | erg/s | kpc | yr | G | G | |
| J0205+6449 | J0205+6449 | 130.72 | 3.08 | 65.727 | $1.90\,10^{-13}$ | $2.64\,10^{37}$ | $2.44\,10^{34}$ | 1.95 | $5.48\,10^{3}$ | $3.58\,10^{12}$ | $1.16\,10^{5}$ | |
| J0248+6021 | J0248+6021 | 136.90 | 0.70 | 217.107 | $5.50\,10^{-14}$ | $2.12\,10^{35}$ | $2.47\,10^{34}$ | 2.0 | $6.25\,10^{4}$ | $3.50\,10^{12}$ | $3.15\,10^{3}$ | |
| J0534+2200 | B0531+21 | 184.56 | -5.78 | 33.635 | $4.20\,10^{-13}$ | $4.36\,10^{38}$ | $6.19\,10^{35}$ | 2.0 | $1.27\,10^{3}$ | $3.80\,10^{12}$ | $9.21\,10^{5}$ | **Crab** |
| J0631+1036 | J0631+1036 | 201.22 | 0.45 | 287.803 | $1.05\,10^{-13}$ | $1.73\,10^{35}$ | $5.57\,10^{33}$ | 1.0 | $4.36\,10^{4}$ | $5.55\,10^{12}$ | $2.14\,10^{3}$ | |
| J0659+1414 | B0656+14 | 201.11 | 8.26 | 384.919 | $5.50\,10^{-14}$ | $3.81\,10^{34}$ | $2.35\,10^{32}$ | 0.28 | $1.11\,10^{5}$ | $4.66\,10^{12}$ | $7.52\,10^{2}$ | |
| J0729-1448 | J0729-1448 | 230.39 | 1.42 | 251.691 | $1.14\,10^{-13}$ | $2.82\,10^{35}$ | $1.72\,10^{34}$ | 3.52 | $3.50\,10^{4}$ | $5.42\,10^{12}$ | $3.13\,10^{3}$ | |
| J0742-2822 | B0740-28 | 243.77 | -2.44 | 166.771 | $1.68\,10^{-14}$ | $1.43\,10^{35}$ | $8.89\,10^{33}$ | 2.07 | $1.57\,10^{5}$ | $1.69\,10^{12}$ | $3.36\,10^{3}$ | |
| J0835-4510 | B0833-45 | 263.55 | -2.79 | 89.365 | $1.25\,10^{-13}$ | $6.90\,10^{36}$ | $8.93\,10^{34}$ | 0.29 | $1.14\,10^{4}$ | $3.38\,10^{12}$ | $4.36\,10^{4}$ | Vela |
| J0908-4913 | B0906-49 | 270.27 | -1.02 | 106.755 | $1.51\,10^{-14}$ | $4.90\,10^{35}$ | $3.50\,10^{34}$ | 2.57 | $1.12\,10^{5}$ | $1.29\,10^{12}$ | $9.72\,10^{3}$ | |
| J0940-5428 | J0940-5428 | 277.51 | -1.29 | 87.545 | $3.28\,10^{-14}$ | $1.93\,10^{36}$ | $4.22\,10^{33}$ | 2.95 | $4.23\,10^{4}$ | $1.71\,10^{12}$ | $2.35\,10^{4}$ | |
| J1016-5857 | J1016-5857 | 284.08 | -1.88 | 107.386 | $8.06\,10^{-14}$ | $2.57\,10^{36}$ | $5.48\,10^{34}$ | 2.9 | $2.11\,10^{4}$ | $2.98\,10^{12}$ | $2.21\,10^{4}$ | |
| J1019-5749 | J1019-5749 | 283.84 | -0.68 | 162.506 | $2.01\,10^{-14}$ | $1.84\,10^{35}$ | $4.27\,10^{34}$ | 6.8 | $1.28\,10^{5}$ | $1.83\,10^{12}$ | $3.92\,10^{3}$ | |
| J1028-5819 | J1028-5819 | 285.07 | -0.50 | 91.403 | $1.61\,10^{-14}$ | $8.33\,10^{35}$ | $1.58\,10^{35}$ | 2.33 | $8.99\,10^{4}$ | $1.23\,10^{12}$ | $1.48\,10^{4}$ | |
| J1048-5832 | B1046-58 | 287.43 | 0.58 | 123.701 | $9.57\,10^{-14}$ | $2.00\,10^{36}$ | $1.76\,10^{35}$ | 2.74 | $2.05\,10^{4}$ | $3.48\,10^{12}$ | $1.69\,10^{4}$ | |
| J1057-5226 | B1055-52 | 285.98 | 6.65 | 197.114 | $5.83\,10^{-15}$ | $3.01\,10^{34}$ | $4.33\,10^{33}$ | 0.35 | $5.35\,10^{5}$ | $1.09\,10^{12}$ | $1.30\,10^{3}$ | Egret |
| J1105-6107 | J1105-6107 | 290.49 | -0.85 | 63.198 | $1.58\,10^{-14}$ | $2.48\,10^{36}$ | $1.45\,10^{35}$ | 4.98 | $6.32\,10^{4}$ | $1.01\,10^{12}$ | $3.69\,10^{4}$ | |
| J1112-6103 | J1112-6103 | 291.22 | -0.46 | 64.962 | $3.15\,10^{-14}$ | $4.54\,10^{36}$ | $3.62\,10^{35}$ | 12.2 | $3.27\,10^{4}$ | $1.45\,10^{12}$ | $4.86\,10^{4}$ | |
| J1119-6127 | J1119-6127 | 292.15 | -0.54 | 408.732 | $4.03\,10^{-12}$ | $2.33\,10^{36}$ | $6.03\,10^{35}$ | 8.4 | $1.61\,10^{3}$ | $4.11\,10^{13}$ | $5.53\,10^{3}$ | |
| J1124-5916 | J1124-5916 | 292.04 | 1.75 | 135.493 | $7.50\,10^{-13}$ | $1.19\,10^{37}$ | $1.70\,10^{35}$ | 4.8 | $2.86\,10^{3}$ | $1.02\,10^{13}$ | $3.78\,10^{4}$ | |
| J1357-6429 | J1357-6429 | 309.92 | -2.51 | 166.167 | $3.57\,10^{-13}$ | $3.07\,10^{36}$ | $2.53\,10^{34}$ | 2.5 | $7.37\,10^{3}$ | $7.80\,10^{12}$ | $1.56\,10^{4}$ | |
| J1410-6132 | J1410-6132 | 312.19 | -0.09 | 50.052 | $3.18\,10^{-14}$ | $1.00\,10^{37}$ | $7.66\,10^{35}$ | 15.6 | $2.50\,10^{4}$ | $1.28\,10^{12}$ | $9.37\,10^{4}$ | |
| J1420-6048 | J1420-6048 | 313.54 | 0.23 | 68.202 | $8.29\,10^{-14}$ | $1.03\,10^{37}$ | $6.39\,10^{35}$ | 5.61 | $1.30\,10^{4}$ | $2.41\,10^{12}$ | $6.98\,10^{4}$ | |
| J1509-5850 | J1509-5850 | 319.97 | -0.62 | 88.925 | $9.17\,10^{-15}$ | $5.15\,10^{35}$ | $1.05\,10^{35}$ | 2.62 | $1.54\,10^{5}$ | $9.14\,10^{11}$ | $1.20\,10^{4}$ | |
| J1513-5908 | B1509-58 | 320.32 | -1.16 | 151.578 | $1.53\,10^{-12}$ | $1.73\,10^{37}$ | $6.88\,10^{34}$ | 4.21 | $1.57\,10^{3}$ | $1.54\,10^{13}$ | $4.07\,10^{4}$ | Comptel |
| J1531-5610 | J1531-5610 | 323.90 | 0.03 | 84.201 | $1.38\,10^{-14}$ | $9.12\,10^{35}$ | $1.01\,10^{33}$ | 2.09 | $9.67\,10^{4}$ | $1.09\,10^{12}$ | $1.68\,10^{4}$ | |
| J1648-4611 | J1648-4611 | 339.44 | -0.79 | 164.958 | $2.37\,10^{-14}$ | $2.09\,10^{35}$ | $1.60\,10^{35}$ | 4.96 | $1.10\,10^{5}$ | $2.00\,10^{12}$ | $4.11\,10^{3}$ | |
| J1702-4128 | J1702-4128 | 344.74 | 0.12 | 182.153 | $5.23\,10^{-14}$ | $3.42\,10^{35}$ | $7.67\,10^{34}$ | 4.75 | $5.52\,10^{4}$ | $3.12\,10^{12}$ | $4.76\,10^{3}$ | |
| J1709-4429 | B1706-44 | 343.11 | -2.67 | 102.496 | $9.28\,10^{-14}$ | $3.40\,10^{36}$ | $8.53\,10^{35}$ | 2.3 | $1.75\,10^{4}$ | $3.12\,10^{12}$ | $2.67\,10^{4}$ | Egret |
| J1718-3825 | J1718-3825 | 348.95 | -0.43 | 74.675 | $1.32\,10^{-14}$ | $1.25\,10^{36}$ | $1.38\,10^{35}$ | 3.6 | $8.98\,10^{4}$ | $1.00\,10^{12}$ | $2.22\,10^{4}$ | |
| J1730-3350 | B1727-33 | 354.14 | 0.09 | 139.497 | $8.48\,10^{-14}$ | $1.23\,10^{36}$ | $3.56\,10^{34}$ | 3.55 | $2.61\,10^{4}$ | $3.48\,10^{12}$ | $1.18\,10^{4}$ | |
| J1741-2054 | J1741-2054 | 6.42 | 4.91 | 413.701 | $1.70\,10^{-14}$ | $9.47\,10^{33}$ | $2.06\,10^{33}$ | 0.38 | $3.86\,10^{5}$ | $2.68\,10^{12}$ | $3.49\,10^{2}$ | |
| J1747-2958 | J1747-2958 | 359.31 | -0.84 | 98.827 | $6.13\,10^{-14}$ | $2.51\,10^{36}$ | $5.70\,10^{35}$ | 4.75 | $2.55\,10^{4}$ | $2.49\,10^{12}$ | $2.38\,10^{4}$ | |
| J1801-2451 | B1757-24 | 5.25 | -0.89 | 124.948 | $1.27\,10^{-13}$ | $2.57\,10^{36}$ | $3.91\,10^{34}$ | 5.22 | $1.56\,10^{4}$ | $4.03\,10^{12}$ | $1.90\,10^{4}$ | |
| J1833-1034 | J1833-1034 | 21.50 | -0.89 | 61.888 | $2.02\,10^{-13}$ | $3.36\,10^{37}$ | $1.56\,10^{35}$ | 4.7 | $4.85\,10^{3}$ | $3.58\,10^{12}$ | $1.39\,10^{5}$ | |
| J1835-1106 | J1835-1106 | 21.22 | -1.51 | 165.916 | $2.06\,10^{-14}$ | $1.78\,10^{35}$ | $5.80\,10^{33}$ | 2.83 | $1.28\,10^{5}$ | $1.87\,10^{12}$ | $3.77\,10^{3}$ | |
| J1952+3252 | B1951+32 | 68.77 | 2.82 | 39.534 | $5.83\,10^{-15}$ | $3.72\,10^{36}$ | $6.61\,10^{34}$ | 2.0 | $1.07\,10^{5}$ | $4.86\,10^{11}$ | $7.24\,10^{4}$ | Egret |
| J2021+3651 | J2021+3651 | 75.22 | 0.11 | 103.742 | $9.56\,10^{-14}$ | $3.38\,10^{36}$ | $5.91\,10^{36}$ | 10.0 | $1.72\,10^{4}$ | $3.19\,10^{12}$ | $2.63\,10^{4}$ | |
| J2030+3641 | J2030+3641 | 76.12 | -1.44 | 200.129 | $6.51\,10^{-15}$ | $3.20\,10^{34}$ | $3.38\,10^{34}$ | 3.0 | $4.87\,10^{5}$ | $1.15\,10^{12}$ | $1.33\,10^{3}$ | |
| J2032+4127 | J2032+4127 | 80.22 | 1.03 | 143.248 | $2.04\,10^{-14}$ | $2.73\,10^{35}$ | $1.69\,10^{35}$ | 3.65 | $1.11\,10^{5}$ | $1.73\,10^{12}$ | $5.41\,10^{3}$ | |
| J2043+2740 | J2043+2740 | 70.61 | -9.15 | 96.131 | $1.23\,10^{-15}$ | $5.46\,10^{34}$ | $3.83\,10^{33}$ | 1.8 | $1.24\,10^{6}$ | $3.48\,10^{11}$ | $3.60\,10^{3}$ | |
| J2229+6114 | J2229+6114 | 106.65 | 2.95 | 51.643 | $7.79\,10^{-14}$ | $2.23\,10^{37}$ | $1.94\,10^{34}$ | 0.8 | $1.05\,10^{4}$ | $2.03\,10^{12}$ | $1.36\,10^{5}$ | |
| J2240+5832 | J2240+5832 | 106.57 | -0.11 | 139.941 | $1.52\,10^{-14}$ | $2.19\,10^{35}$ | $7.68\,10^{34}$ | 7.7 | $1.46\,10^{5}$ | $1.48\,10^{12}$ | $4.96\,10^{3}$ | |

| TABLE 2 - RADIO QUIET INSs | | | | | | | | | | | | |
|---|---|---|---|---|---|---|---|---|---|---|---|---|
| JName | BName | Glat | Glon | P | Pdot | Edot | Lum | Dist | Age | Bs | B_LC | Comments |
| | | deg | deg | ms | s/s | erg/s | erg/s | kpc | yr | G | G | |
| J0007+7303 | J0007+7303 | 119.66 | 10.46 | 315.893 | $3.57\;10^{-13}$ | $4.48\;10^{35}$ | $9.39\;10^{34}$ | 1.4 | $1.40\;10^{4}$ | $1.08\;10^{13}$ | $3.14\;10^{3}$ | CTA-1 |
| J0106+4855 | J0106+4855 | 125.47 | -13.87 | 83.157 | $4.28\;10^{-16}$ | $2.94\;10^{34}$ | $2.09\;10^{34}$ | 3.01 | $3.08\;10^{6}$ | $1.91\;10^{11}$ | $3.06\;10^{3}$ | |
| J0357+3205 | J0357+3205 | 162.76 | -16.01 | 444.105 | $1.31\;10^{-14}$ | $5.90\;10^{33}$ | $<5.16\;10^{35}$ | <8.2 | $5.37\;10^{5}$ | $2.44\;10^{12}$ | $2.56\;10^{2}$ | |
| J0622+3749 | J0622+3749 | 175.88 | 10.96 | 333.208 | $2.54\;10^{-14}$ | $2.71\;10^{34}$ | $<1.19\;10^{35}$ | <8.3 | $2.08\;10^{5}$ | $2.94\;10^{12}$ | $7.33\;10^{2}$ | |
| J0633+0632 | J0633+0632 | 205.09 | -0.93 | 297.397 | $7.96\;10^{-14}$ | $1.19\;10^{35}$ | $<8.52\;10^{35}$ | <8.7 | $5.92\;10^{4}$ | $4.92\;10^{12}$ | $1.72\;10^{3}$ | |
| J0633+1746 | J0633+1746 | 195.13 | 4.27 | 237.104 | $1.10\;10^{-14}$ | $3.25\;10^{34}$ | $3.17\;10^{34}$ | 0.25 | $3.42\;10^{5}$ | $1.63\;10^{12}$ | $1.13\;10^{3}$ | **Geminga** |
| J0734-1559 | J0734-1559 | 232.06 | 2.02 | 155.141 | $1.25\;10^{-14}$ | $1.32\;10^{35}$ | $<7.05\;10^{35}$ | <10.3 | $1.96\;10^{5}$ | $1.41\;10^{12}$ | $3.48\;10^{3}$ | |
| J1023-5746 | J1023-5746 | 284.17 | -0.41 | 111.479 | $3.82\;10^{-13}$ | $1.09\;10^{37}$ | $<6.58\;10^{36}$ | <16.8 | $4.62\;10^{3}$ | $6.61\;10^{12}$ | $4.39\;10^{4}$ | |
| J1044-5737 | J1044-5737 | 286.58 | 1.16 | 139.030 | $5.46\;10^{-14}$ | $8.02\;10^{35}$ | $<5.52\;10^{36}$ | <17.2 | $4.03\;10^{4}$ | $2.79\;10^{12}$ | $9.55\;10^{3}$ | |
| J1135-6055 | J1135-6055 | 293.79 | 0.58 | 114.487 | $7.84\;10^{-14}$ | $2.06\;10^{36}$ | $<1.94\;10^{36}$ | <18.4 | $2.31\;10^{4}$ | $3.03\;10^{12}$ | $1.86\;10^{4}$ | |
| J1413-6205 | J1413-6205 | 312.37 | -0.74 | 109.741 | $2.74\;10^{-14}$ | $8.18\;10^{35}$ | $<8.62\;10^{36}$ | <21.4 | $6.35\;10^{4}$ | $1.75\;10^{12}$ | $1.22\;10^{4}$ | |
| J1418-6058 | J1418-6058 | 313.32 | 0.13 | 110.577 | $1.69\;10^{-13}$ | $4.94\;10^{36}$ | $9.24\;10^{34}$ | 1.6 | $1.03\;10^{4}$ | $4.38\;10^{12}$ | $2.98\;10^{4}$ | |
| J1429-5911 | J1429-5911 | 315.26 | 1.30 | 115.844 | $3.05\;10^{-14}$ | $7.74\;10^{35}$ | $<4.57\;10^{36}$ | <21.8 | $6.02\;10^{4}$ | $1.90\;10^{12}$ | $1.13\;10^{4}$ | |
| J1459-6053 | J1459-6053 | 317.89 | -1.79 | 103.151 | $2.53\;10^{-14}$ | $9.09\;10^{35}$ | $<7.62\;10^{36}$ | <22.2 | $6.47\;10^{4}$ | $1.63\;10^{12}$ | $1.37\;10^{4}$ | |
| J1620-4927 | J1620-4927 | 333.89 | 0.41 | 171.935 | $1.05\;10^{-14}$ | $8.15\;10^{34}$ | $<1.08\;10^{37}$ | <24.1 | $2.60\;10^{5}$ | $1.36\;10^{12}$ | $2.46\;10^{3}$ | |
| J1732-3131 | J1732-3131 | 356.31 | 1.01 | 196.544 | $2.80\;10^{-14}$ | $1.46\;10^{35}$ | $8.62\;10^{33}$ | 0.61 | $1.11\;10^{5}$ | $2.38\;10^{12}$ | $2.88\;10^{3}$ | |
| J1746-3239 | J1746-3239 | 356.96 | -2.17 | 199.541 | $6.56\;10^{-15}$ | $3.26\;10^{34}$ | $<5.53\;10^{36}$ | <25.3 | $4.82\;10^{5}$ | $1.16\;10^{12}$ | $1.34\;10^{3}$ | |
| J1803-2149 | J1803-2149 | 8.14 | 0.19 | 106.332 | $1.95\;10^{-14}$ | $6.41\;10^{35}$ | $<7.01\;10^{36}$ | <25.2 | $8.63\;10^{4}$ | $1.46\;10^{12}$ | $1.12\;10^{4}$ | |
| J1809-2332 | J1809-2332 | 7.39 | -1.99 | 146.789 | $3.44\;10^{-14}$ | $4.30\;10^{35}$ | $1.64\;10^{35}$ | 1.7 | $6.76\;10^{4}$ | $2.27\;10^{12}$ | $6.62\;10^{3}$ | Taz |
| J1813-1246 | J1813-1246 | 17.24 | 2.44 | 48.073 | $1.76\;10^{-14}$ | $6.24\;10^{36}$ | $<1.84\;10^{37}$ | <24.7 | $4.34\;10^{4}$ | $9.30\;10^{11}$ | $7.70\;10^{4}$ | |
| J1826-1256 | J1826-1256 | 18.56 | -0.38 | 110.227 | $1.21\;10^{-13}$ | $3.58\;10^{36}$ | $<2.79\;10^{37}$ | <24.7 | $1.44\;10^{4}$ | $3.70\;10^{12}$ | $2.54\;10^{4}$ | Eel |
| J1836+5925 | J1836+5925 | 88.88 | 25.00 | 173.264 | $1.50\;10^{-15}$ | $1.14\;10^{34}$ | $2.04\;10^{34}$ | 0.53 | $1.83\;10^{6}$ | $5.16\;10^{11}$ | $9.14\;10^{2}$ | Next Geminga |
| J1838-0537 | J1838-0537 | 26.51 | 0.21 | 145.709 | $4.65\;10^{-13}$ | $5.93\;10^{36}$ | $<1.30\;10^{37}$ | <24.1 | $4.97\;10^{3}$ | $8.33\;10^{12}$ | $2.48\;10^{4}$ | |
| J1846+0919 | J1846+0919 | 40.69 | 5.34 | 225.552 | $9.93\;10^{-15}$ | $3.42\;10^{34}$ | $<1.40\;10^{36}$ | <22.0 | $3.60\;10^{5}$ | $1.51\;10^{12}$ | $1.21\;10^{3}$ | |
| J1907+0602 | J1907+0602 | 40.18 | -0.89 | 106.636 | $8.67\;10^{-14}$ | $2.82\;10^{36}$ | $3.14\;10^{35}$ | 3.21 | $1.95\;10^{4}$ | $3.08\;10^{12}$ | $2.34\;10^{4}$ | |
| J1954+2836 | J1954+2836 | 65.24 | 0.38 | 92.710 | $2.12\;10^{-14}$ | $1.05\;10^{36}$ | $<4.28\;10^{36}$ | <18.6 | $6.94\;10^{4}$ | $1.42\;10^{12}$ | $1.64\;10^{4}$ | |
| J1957+5033 | J1957+5033 | 84.60 | 11.00 | 374.806 | $6.83\;10^{-15}$ | $5.12\;10^{33}$ | $<6.59\;10^{35}$ | <14.5 | $8.69\;10^{5}$ | $1.62\;10^{12}$ | $2.83\;10^{2}$ | |
| J1958+2846 | J1958+2846 | 65.88 | -0.35 | 290.397 | $2.12\;10^{-13}$ | $3.42\;10^{35}$ | $<3.74\;10^{36}$ | <18.5 | $2.17\;10^{4}$ | $7.94\;10^{12}$ | $2.98\;10^{3}$ | |
| J2021+4026 | J2021+4026 | 78.23 | 2.09 | 265.320 | $5.42\;10^{-14}$ | $1.14\;10^{35}$ | $2.57\;10^{35}$ | 1.5 | $7.76\;10^{4}$ | $3.84\;10^{12}$ | $1.89\;10^{3}$ | γ Cygni |
| J2028+3332 | J2028+3332 | 73.36 | -3.01 | 176.707 | $4.86\;10^{-15}$ | $3.48\;10^{34}$ | $<2.06\;10^{36}$ | <17.2 | $5.76\;10^{5}$ | $9.37\;10^{11}$ | $1.56\;10^{3}$ | |
| J2030+4415 | J2030+4415 | 82.34 | 2.89 | 227.070 | $6.49\;10^{-15}$ | $2.19\;10^{34}$ | $<1.69\;10^{36}$ | <15.7 | $5.54\;10^{5}$ | $1.23\;10^{12}$ | $9.66\;10^{2}$ | |
| J2055+2539 | J2055+2539 | 70.69 | -12.52 | 319.561 | $4.11\;10^{-15}$ | $4.97\;10^{33}$ | $<1.50\;10^{36}$ | <15.3 | $1.23\;10^{6}$ | $1.16\;10^{12}$ | $3.27\;10^{2}$ | |
| J2111+4606 | J2111+4606 | 88.31 | -1.45 | 157.830 | $1.43\;10^{-13}$ | $1.44\;10^{36}$ | $<1.15\;10^{36}$ | <14.8 | $1.75\;10^{4}$ | $4.81\;10^{12}$ | $1.13\;10^{4}$ | |
| J2139+4716 | J2139+4716 | 92.63 | -4.02 | 282.849 | $1.80\;10^{-15}$ | $3.15\;10^{33}$ | $<5.56\;10^{35}$ | <14.1 | $2.49\;10^{6}$ | $7.23\;10^{11}$ | $2.94\;10^{2}$ | |
| J2238+5903 | J2238+5903 | 106.56 | 0.48 | 162.734 | $9.70\;10^{-14}$ | $8.88\;10^{35}$ | $<1.18\;10^{36}$ | <12.4 | $2.66\;10^{4}$ | $4.02\;10^{12}$ | $8.59\;10^{3}$ | |

| TABLE 3 - MSPs | | | | | | | | | | | | |
|---|---|---|---|---|---|---|---|---|---|---|---|---|
| JName | BName | Glat | Glon | P | Pdot | Edot | Lum | Dist | Age | Bs | B_LC | Comments |
| | | deg | deg | ms | s/s | erg/s | erg/s | kpc | yr | G | G | |
| J0023+0923 | J0023+0923 | 111.38 | -52.85 | 3.050 | $1.09\,10^{-20}$ | $1.51\,10^{34}$ | $4.56\,10^{32}$ | 0.69 | $4.45\,10^{09}$ | $1.84\,10^{08}$ | $5.97\,10^{04}$ | |
| J0030+0451 | J0030+0451 | 113.14 | -57.61 | 4.870 | $1.02\,10^{-20}$ | $3.62\,10^{33}$ | $5.75\,10^{32}$ | 0.28 | $7.28\,10^{09}$ | $2.30\,10^{08}$ | $1.83\,10^{04}$ | |
| J0034-0534 | J0034-0534 | 111.49 | -68.07 | 1.880 | $4.98\,10^{-21}$ | $1.72\,10^{34}$ | $5.67\,10^{32}$ | 0.54 | $1.03\,10^{10}$ | $7.46\,10^{07}$ | $1.03\,10^{05}$ | |
| J0101-6422 | J0101-6422 | 301.19 | -52.72 | 2.570 | $4.80\,10^{-21}$ | $1.01\,10^{34}$ | $3.79\,10^{32}$ | 0.55 | $9.42\,10^{09}$ | $1.07\,10^{08}$ | $5.79\,10^{04}$ | |
| J0102+4839 | J0102+4839 | 124.87 | -14.17 | 2.960 | $1.17\,10^{-20}$ | $1.75\,10^{34}$ | $8.51\,10^{33}$ | 2.32 | $4.01\,10^{09}$ | $1.88\,10^{08}$ | $6.68\,10^{04}$ | |
| J0218+4232 | J0218+4232 | 139.51 | -17.53 | 2.320 | $7.74\,10^{-20}$ | $2.43\,10^{35}$ | $3.80\,10^{34}$ | 2.64 | $4.78\,10^{08}$ | $4.27\,10^{08}$ | $3.15\,10^{05}$ | |
| J0340+4130 | J0340+4130 | 153.78 | -11.02 | 3.300 | $5.90\,10^{-21}$ | $7.87\,10^{33}$ | $7.29\,10^{33}$ | 1.73 | $8.86\,10^{09}$ | $1.41\,10^{08}$ | $3.62\,10^{04}$ | |
| J0437-4715 | J0437-4715 | 253.39 | -41.96 | 5.760 | $5.73\,10^{-20}$ | $2.91\,10^{33}$ | $4.86\,10^{31}$ | 0.16 | $6.49\,10^{09}$ | $2.88\,10^{08}$ | $1.39\,10^{04}$ | |
| J0610-2100 | J0610-2100 | 227.75 | -18.18 | 3.860 | $1.23\,10^{-20}$ | $8.35\,10^{32}$ | $9.85\,10^{33}$ | 3.54 | $5.03\,10^{10}$ | $6.93\,10^{07}$ | $1.11\,10^{04}$ | |
| J0613-0200 | J0613-0200 | 210.41 | -9.30 | 3.060 | $9.59\,10^{-21}$ | $1.20\,10^{34}$ | $2.90\,10^{33}$ | 0.9 | $5.55\,10^{09}$ | $1.65\,10^{08}$ | $5.31\,10^{04}$ | |
| J0614-3329 | J0614-3329 | 240.50 | -21.83 | 3.150 | $1.78\,10^{-20}$ | $2.20\,10^{34}$ | $4.72\,10^{34}$ | 1.9 | $2.80\,10^{09}$ | $2.40\,10^{08}$ | $7.06\,10^{04}$ | |
| J0751+1807 | J0751+1807 | 202.73 | 21.09 | 3.480 | $7.78\,10^{-21}$ | $7.21\,10^{33}$ | $2.54\,10^{32}$ | 0.4 | $7.16\,10^{09}$ | $1.66\,10^{08}$ | $3.62\,10^{04}$ | |
| J1024-0719 | J1024-0719 | 251.70 | 40.52 | 5.160 | $1.85\,10^{-20}$ | $4.60\,10^{32}$ | $5.67\,10^{31}$ | 0.39 | $5.10\,10^{10}$ | $9.20\,10^{07}$ | $6.17\,10^{03}$ | |
| J1124-3653 | J1124-3653 | 284.10 | 22.76 | 2.410 | $5.75\,10^{-21}$ | $1.71\,10^{34}$ | $4.29\,10^{33}$ | 1.72 | $6.64\,10^{09}$ | $1.19\,10^{08}$ | $7.83\,10^{04}$ | |
| J1125-5825 | J1125-5825 | 291.89 | 2.60 | 3.100 | $6.09\,10^{-20}$ | $8.05\,10^{34}$ | $7.31\,10^{33}$ | 2.62 | $8.06\,10^{08}$ | $4.40\,10^{08}$ | $1.36\,10^{05}$ | |
| J1231-1411 | J1231-1411 | 295.53 | 48.39 | 3.680 | $2.12\,10^{-20}$ | $5.15\,10^{33}$ | $2.36\,10^{33}$ | 0.43 | $8.98\,10^{09}$ | $1.56\,10^{08}$ | $2.89\,10^{04}$ | |
| J1446-4701 | J1446-4701 | 322.50 | 11.43 | 2.190 | $9.85\,10^{-21}$ | $3.68\,10^{34}$ | $1.89\,10^{33}$ | 1.46 | $3.52\,10^{09}$ | $1.49\,10^{08}$ | $1.30\,10^{05}$ | |
| J1514-4946 | J1514-4946 | 325.25 | 6.81 | 3.590 | $1.87\,10^{-20}$ | $1.60\,10^{34}$ | $4.76\,10^{33}$ | 0.94 | $3.04\,10^{09}$ | $2.62\,10^{08}$ | $5.22\,10^{04}$ | |
| J1600-3053 | J1600-3053 | 344.09 | 16.45 | 3.600 | $9.50\,10^{-21}$ | $7.30\,10^{33}$ | $1.68\,10^{33}$ | 1.63 | $6.61\,10^{09}$ | $1.78\,10^{08}$ | $3.52\,10^{04}$ | |
| J1614-2230 | J1614-2230 | 352.64 | 20.19 | 3.150 | $9.62\,10^{-21}$ | $6.33\,10^{33}$ | $1.23\,10^{33}$ | 0.65 | $9.96\,10^{09}$ | $1.27\,10^{08}$ | $3.74\,10^{04}$ | |
| J1658-5324 | J1658-5324 | 334.87 | -6.63 | 2.440 | $1.10\,10^{-20}$ | $3.02\,10^{34}$ | $2.99\,10^{33}$ | 0.93 | $3.51\,10^{09}$ | $1.66\,10^{08}$ | $1.05\,10^{05}$ | |
| J1713+0747 | J1713+0747 | 28.75 | 25.22 | 4.570 | $8.53\,10^{-21}$ | $3.44\,10^{33}$ | $1.34\,10^{33}$ | 1.05 | $8.72\,10^{09}$ | $1.97\,10^{08}$ | $1.90\,10^{04}$ | |
| J1741+1351 | J1741+1351 | 37.89 | 21.64 | 3.750 | $3.02\,10^{-20}$ | $2.18\,10^{34}$ | $3.36\,10^{32}$ | 1.08 | $2.04\,10^{09}$ | $3.34\,10^{08}$ | $5.83\,10^{04}$ | |
| J1744-1134 | J1744-1134 | 14.79 | 9.18 | 4.070 | $8.92\,10^{-21}$ | $4.11\,10^{33}$ | $6.76\,10^{32}$ | 0.41 | $9.19\,10^{09}$ | $1.71\,10^{08}$ | $2.33\,10^{04}$ | |
| J1747-4036 | J1747-4036 | 350.21 | -6.41 | 1.650 | $1.33\,10^{-20}$ | $1.17\,10^{35}$ | $1.36\,10^{34}$ | 3.39 | $1.97\,10^{09}$ | $1.50\,10^{08}$ | $3.07\,10^{05}$ | |
| J1810+1744 | J1810+1744 | 44.64 | 16.81 | 1.660 | $4.63\,10^{-21}$ | $3.97\,10^{34}$ | $1.12\,10^{34}$ | 2.0 | $5.68\,10^{09}$ | $8.87\,10^{07}$ | $1.78\,10^{05}$ | |
| J1823-3021A | B1820-30A | 2.79 | -7.91 | 5.440 | $3.38\,10^{-18}$ | $8.28\,10^{35}$ | $7.39\,10^{34}$ | 7.6 | $2.55\,10^{07}$ | $4.34\,10^{09}$ | $2.48\,10^{05}$ | |
| J1858-2216 | J1858-2216 | 13.58 | -11.39 | 2.380 | $3.87\,10^{-21}$ | $1.13\,10^{34}$ | $7.64\,10^{32}$ | 0.94 | $9.74\,10^{09}$ | $9.71\,10^{07}$ | $6.63\,10^{04}$ | |
| J1902-5105 | J1902-5105 | 345.65 | -22.38 | 1.740 | $9.00\,10^{-21}$ | $6.86\,10^{34}$ | $3.59\,10^{33}$ | 1.18 | $3.06\,10^{09}$ | $1.27\,10^{08}$ | $2.21\,10^{05}$ | |
| J1939+2134 | B1937+21 | 57.51 | -0.29 | 1.560 | $1.05\,10^{-19}$ | $1.10\,10^{36}$ | $1.39\,10^{34}$ | 3.56 | $2.34\,10^{08}$ | $4.10\,10^{08}$ | $9.95\,10^{05}$ | |
| J1959+2048 | B1957+20 | 59.20 | -4.70 | 1.610 | $1.68\,10^{-20}$ | $7.63\,10^{34}$ | $1.26\,10^{34}$ | 2.49 | $3.16\,10^{09}$ | $1.15\,10^{08}$ | $2.54\,10^{05}$ | |
| J2017+0603 | J2017+0603 | 48.62 | -16.03 | 2.900 | $8.30\,10^{-21}$ | $1.30\,10^{34}$ | $9.82\,10^{33}$ | 1.57 | $5.54\,10^{09}$ | $1.57\,10^{08}$ | $5.93\,10^{04}$ | |
| J2043+1711 | J2043+1711 | 61.92 | -15.31 | 2.380 | $5.70\,10^{-21}$ | $1.27\,10^{34}$ | $1.00\,10^{34}$ | 1.76 | $8.73\,10^{09}$ | $1.03\,10^{08}$ | $7.01\,10^{04}$ | |
| J2047+1053 | J2047+1053 | 57.05 | -19.68 | 4.290 | $2.10\,10^{-20}$ | $1.05\,10^{34}$ | $3.10\,10^{33}$ | 2.05 | $3.24\,10^{09}$ | $3.04\,10^{08}$ | $3.54\,10^{04}$ | |
| J2051-0827 | J2051-0827 | 39.19 | -30.41 | 4.510 | $1.28\,10^{-20}$ | $5.42\,10^{33}$ | $4.37\,10^{32}$ | 1.04 | $5.68\,10^{09}$ | $2.41\,10^{08}$ | $2.42\,10^{04}$ | |
| J2124-3358 | J2124-3358 | 10.93 | -45.44 | 4.930 | $2.06\,10^{-20}$ | $3.67\,10^{33}$ | $3.96\,10^{32}$ | 0.3 | $7.01\,10^{09}$ | $2.37\,10^{08}$ | $1.82\,10^{04}$ | |
| J2214+3000 | J2214+3000 | 86.86 | -21.67 | 3.120 | $1.50\,10^{-20}$ | $1.92\,10^{34}$ | $9.29\,10^{33}$ | 1.54 | $3.30\,10^{09}$ | $2.19\,10^{08}$ | $6.63\,10^{04}$ | |
| J2215+5135 | J2215+5135 | 99.87 | -4.16 | 2.610 | $2.34\,10^{-20}$ | $5.19\,10^{34}$ | $1.28\,10^{34}$ | 3.01 | $1.77\,10^{09}$ | $2.50\,10^{08}$ | $1.29\,10^{05}$ | |
| J2241-5236 | J2241-5236 | 337.46 | -54.93 | 2.190 | $8.70\,10^{-21}$ | $2.60\,10^{34}$ | $1.05\,10^{33}$ | 0.51 | $3.99\,10^{09}$ | $1.40\,10^{08}$ | $1.22\,10^{05}$ | |
| J2302+4442 | J2302+4442 | 103.40 | -14.00 | 5.190 | $1.33\,10^{-20}$ | $3.82\,10^{33}$ | $6.21\,10^{33}$ | 1.19 | $6.18\,10^{09}$ | $2.66\,10^{08}$ | $1.75\,10^{04}$ | |